\documentclass{emulateapj}
\usepackage{graphicx, epsfig}

\begin{document}

\title{Beyond Caustic Crossings: Properties of Binary Microlensing
Light Curves}
\author{Christopher Night \altaffilmark{1}}
\author{Rosanne Di Stefano \altaffilmark{1,2,3}}
\author{Megan Schwamb \altaffilmark{4}}
\altaffiltext{1}{Harvard-Smithsonian Center for Astrophysics, 60
Garden Street, Cambridge, MA 02138}
\altaffiltext{2}{Kavli Institute for Theoretical Physics, University  
of California Santa Barbara, Santa Barbara, CA 93106}
\altaffiltext{3}{Department of Physics and Astronomy, Tufts
University, Medford, MA 02155}
\altaffiltext{4}{California Institute of Technology, Pasadena, CA}
\submitted{Submitted to ApJ, 23 April 2005} 

\begin{abstract}
Binary microlensing light curves have a variety of morphologies. Many
are indistinguishable from point lens light curves. Of those that
deviate from the point lens form, caustic crossing light curves have
tended to dominate identified binary lens events. Other distinctive
signatures of binary lenses include significant asymmetry, multiple
peaks, and repeating events. We have quantified, using high resolution
simulations, the theoretically expected relative numbers of each type
of binary lens event, based on its measurable characteristics. We find
that a microlensing survey with current levels of photometric
uncertainty and sampling should find at least as many non-caustic
crossing binary lens events as caustic crossing events; in future
surveys with more sensitive photometry, the contribution of
distinctive non-caustic crossing events will be even greater. To try
to explain why caustic crossing light curves appear to be so dominant
among the published binary lensing events, we investigate the
influence of several physical effects, including blending, sampling
rate, and various binary populations.
\end{abstract}

\section{Introduction} 
\label{introduction}

\subsection{Background} 
\label{background}

In gravitational microlensing, a lens and a more distant source pass
near each other in the sky, causing an apparent brightening (or
magnification) of the source that varies with time as the alignment
changes. In the simplest case, when the source is a point source and
the lens is a point mass, the light curves associated with these
events are described by a simple formula
\citep{Einstein1936}. Relative to this point lens form, the light
curves produced by a binary lens, such as a binary star, are much more
diverse.

General relativity is a non-linear theory, and binary lens light
curves can therefore differ significantly from the sum of their two
point lens components. Furthermore, although the shape of the light
curve associated with any given event can be computed to arbitrary
accuracy, there is no simple analytic formula for binary lens light
curves. These light curves can be practically indistinguishable from
point lens light curves, or they can be distinctive, as in the case of
caustic crossing events, which exhibit spikes of divergent
magnification. Other features not found in point lens light curves
include asymmetry in time, multiple local maxima, and repeated events,
which comprise two distinct brightenings separated by a return to
baseline.

The distinctive events published to date by surveys such as MACHO and
OGLE have tended to be dominated by caustic crossing events. The MACHO
data contain 21 known binary lensing events, of which 14 were caustic
crossing; toward the bulge, 12 out of 16 binary events showed caustic
crossings \citep{MACHOBinaries}. The OGLE III 2002-2003 season
produced 24 identified binary lensing events, of which 17 were
caustic crossing \citep{OGLEBinaries2002}, and the 2004 season
produced 25 events, of which 22 were caustic crossing
\citep{OGLEBinaries2004}.

\subsection{Smoothly perturbed events}
\label{smoothly-perturbed-events}

Our main focus is on binary lensing events that differ significantly
from the point lens form, but that do not exhibit caustic
crossings. We call these events {\it smoothly perturbed} if they differ
significantly from the best fit point lens model. The amount by which
they must differ from the point lens form in order to be considered
perturbed depends on the photometric uncertainty of a survey. Our main
question is: for typical values of photometric uncertainty, how many
such smoothly perturbed events are expected for each detected caustic
crossing event?  The answer to this and similar questions depends on
the characteristics of the binary lens, specifically the mass ratio $q$
and the orbital separation $a$.

In \S 2, we introduce the binary lensing parameters, and explore the
dependence of the rates of different types of light curves on these
parameters.  In \S 3 we describe the simulations we have used to
calculate these relative rates numerically. In \S 4 we give the
results for individual binary lenses over the ranges of our
simulation, as characterized by a set of binary parameters. In \S 5 we
give the results for typical populations of binary lenses and study
the implications. We find that current surveys should identify
significantly more smoothly perturbed binary lens event than surveys
have until now.

\section{Lensing geometry}
\label{lensing-geometry}

\subsection{Lensing formula}
\label{lensing-formula}

The magnification associated with a binary lens depends on the binary
parameters and the relative locations of the source and of the lens
components on the plane of the sky.  As the relative positions change,
we can compute the magnification at each time \citep{MaoPaczynski,
SchneidersBook, PettersBook}.

\begin{figure*}[th]
\begin{center}
\includegraphics[width=4.5in]{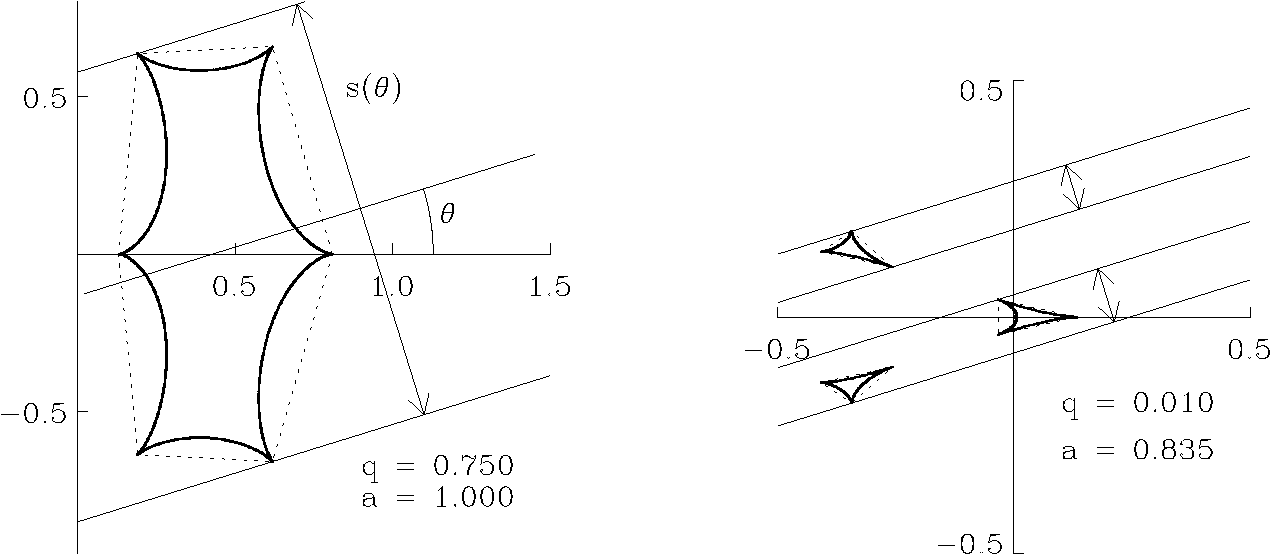}
\end{center}
\caption{
Graphical depiction of $\langle s \rangle$ for caustic crossing events
for two specific binaries.  The left panel shows the binary
$(q,a)=(0.750,1.000)$, which has a single connected caustic. The
$x$-$y$ plane shown is the lens plane, with lengths measured in units of
the Einstein angle $\theta_E$. For a given value of $\theta$,
$s(\theta )$ is the width of the caustic region, as seen from that
angle. The minimum bounding polygon (dashed) has a perimeter equal to
$\pi$ times the average value of $s(\theta )$, that is, $\pi \langle
s\rangle$. Since this polygon is defined by the positions of the cusps
at the corners, we can calculate $\langle s \rangle$ from the cusp
positions alone, and check the result against the simulation. The
right panel shows the binary $(q,a)=(0.010,0.835)$, for which there
are three disconnected caustics. In this case, $s(\theta )$ may
be the sum of two or three widths. The diagram shows $s(\theta )$ for
the same value of $\theta$ as the left diagram. Again it is possible
to calculate $\langle s \rangle$ from the cusp positions alone. }
\label{geometry-fig}
\end{figure*}

It is convenient to express the angular positions and angular
distances involved in units of the Einstein angle $\theta_E =
\sqrt{\frac{4G M}{c^2}\frac{D_S-D_L}{D_L D_S}}$, where $M$ is the
total mass of the lens, $D_S$ is the distance to the source, and $D_L$
is the distance to the lens. For a given $M$, the lens is described
by two parameters, the mass ratio $q=M_2/M_1$ and the instantaneous
angular separation $a$, which is measured in units of
$\theta_E$. $M_1$ and $M_2$ are the mass of the primary and the
secondary, and $M = M_1 + M_2$.

To define the coordinate system, we place the primary at the origin
and the secondary on the positive $x$-axis. A point source is
completely described by its position $(x_s, y_s)$ in this coordinate
system.

The formula for the magnification, $A(t)$, is simply a function of the
binary parameters $(q,a)$ and the angular source position
$(x_s(t),y_s(t))$: $A(q,a,x_s(t),y_s(t))$. We use as the units of time
the Einstein angle crossing time $t_E=\theta_E/\omega$, where $\omega$
is the angular speed of the source relative to the fixed lens.

We consider only the case where the lens is static, so neither $a$ nor
the binary's orientation changes. The angular rotation of a binary
lens over the course of an event is expected to be small.  This is
because, although binaries with a wide range of separations may act as
lenses, the binary nature is only evident when the projected
separation between the components is on the order of $\theta_E,$ which
is typically larger than an AU for stellar lenses.  The orbital
periods therefore tend to be longer than the event durations, which
are on the order of weeks to months. \citet{Dominik1998} gives more
detailed criteria for when binary rotation is likely to be
significant.


Thus we consider the source trajectory to be a straight line in the
lens plane, defined by $b$, the angular distance of closest approach
to the lens center of mass, and $\theta$, the angle the trajectory
makes with the binary axis (defined to be the $x$-axis):

\begin{eqnarray}
x_s(t) & = & -b \sin \theta + t \cos \theta + a_1 \\
y_s(t) & = & b \cos \theta + t \sin \theta
\end{eqnarray}

Here $a_1 = a M_2 / M$ is the $x$-coordinate of the lens center of
mass. Symmetry with respect to the binary axis, and symmetry with
respect to time reversal, allow us to cover the complete parameter
space by considering values of $\theta$ in the interval $(0,\pi/2)$.

\subsection{The relative rate measure $\langle s\rangle$}
\label{relative-rate-measure}

We consider a light curve to contain an event if its maximum
magnification is greater than a certain value, $A_{\rm cut}$.  The rate
for lensing events is given by $\sigma\, s \, \omega$, where $\sigma$
is the source density, and $s$ is the width of the lensing region,
that is, the region bounded by the isomagnification contour in the
lens plane corresponding to the cutoff magnification.  Then for a
point lens, with $A_{\rm cut} = 1.34,$ the lensing region is a circle of
radius $\theta_E$, and so $s = 2\theta_E$. For a given binary lens,
the lensing region is not circular, so $s$ depends on the angle of
approach $\theta$.  $s(\theta)$ is the linear size of the lensing
region as seen from the angle $\theta$. The average value of
$s(\theta)$, $\langle s \rangle,$ for a given binary represents the
lensing region's average angular width. If a large number of sources
were to pass behind this binary, the event rate would be $\sigma\,
{\langle s \rangle}\, \omega$.  Since the only dependence on the
intrinsic properties of the lens is through the factor ${\langle s
\rangle}$, it is this quantity we will compute for different lenses.
Taking $\sigma$ and $\omega$ to be the same for all potential lenses,
the {\em ratio} of rates does not depend on them.  We can therefore
use the ratios of the values of $\langle s \rangle$ to compare
relative average rates for the total numbers of events generated by
different lenses.

For a given binary lens, different source paths produce light curves
with different characteristics.  For example, paths that cross the
caustics produce light curves with distinctive wall-like structures,
the so-called caustic crossing light curves. The rate of such events
is determined by the dimensions (in angular units in the lens plane)
of the caustics (see Figure \ref{geometry-fig}).  The value
$s_{cc}(\theta)$ measures the extent of the caustics perpendicular to
the direction of approach specified by $\theta$.

\begin{figure*}[ht]
\begin{center}
$ \begin{array}{cc}
\includegraphics[width=3in]{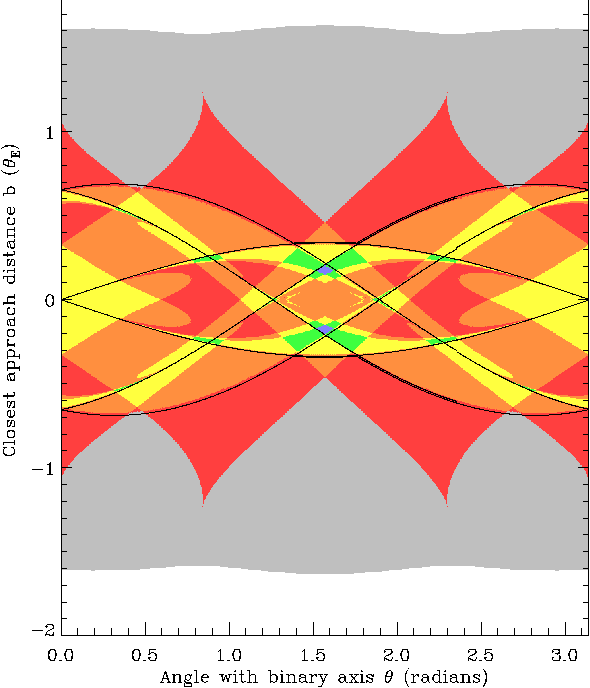} &
\includegraphics[width=3in]{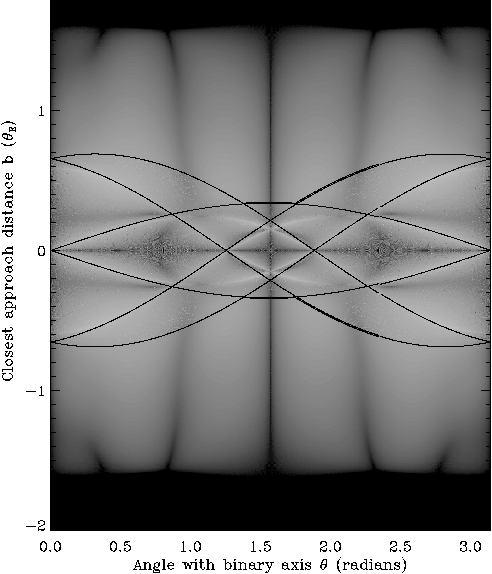}
\end{array} $
\end{center}
\caption{
Peak count and asymmetry parameter in the light curve corresponding to
a given $b$ and $\theta$, with $(q,a) = (1.0,1.0)$ held fixed. The
graph is {\it not} shown in $x$-$y$ coordinates like in Figure
\ref{geometry-fig}; $b$ and $\theta$ are similar to the polar coordinates
in that plane. The left panel shows the number of peaks color
coded. Colors gray, red, yellow, orange, green, and blue correspond to
1, 2, 3, 4, 5, and 6 peaks, respectively. The right panel shows the
asymmetry parameter $k$ (see \S \ref{asymmetry-parameter}), color
coded logarithmically, with black the most symmetric ($k \approx 0$)
and white the least symmetric ($k \approx 1$).  In both panels, black
sinusoidal curves show the $b$-value of the caustic cusps as a
function of $\theta$; any trajectory with values of $b$ and $\theta$
between these black curves will produce a caustic crossing light
curve. The black curves are easy to describe in terms of the cusp
locations in the $x$-$y$ plane (see Figure \ref{geometry-fig}), and so
$\langle s\rangle$ for caustic crossings may be computed from them.
For other light curve characteristics, however, $\langle s\rangle$ is
more complicated. For instance, $\langle s\rangle$ for single-peaked
events is given by $\frac{1}{\pi}$ times the area of the gray region
in the left panel. That is, $\langle s\rangle =
\frac{1}{\pi} \int_0^\pi \int_{-\infty}^{+\infty} N(b,\theta)\,db\,d\theta$,
where $N(b,\theta)$ is 1 at any point in the gray region and 0
elsewhere. In the right panel, $\langle s\rangle$ for light curves
with a certain amount of asymmetry corresponds to the size of a
contour for a certain brightness.  Any other $\langle s\rangle$ is
similarly defined in terms of a region on the $b-\theta$ graph. }
\label{btheta}
\end{figure*}

Even event types that do not correspond to crossing an easily defined
region, such as smoothly perturbed events, may be thought of as having
an effective width, which $s(\theta)$ measures.  For a given $q$, $a$,
and $\theta$, there is a range of values of $b$ that correspond to
light curves with any property we choose to investigate, such as being
smoothly perturbed.  This range may be connected or disconnected, but
generally, the bigger the size of the range (or the sum of the sizes
of the parts), the more likely that a randomly chosen $b$ will
correspond to a light curve with the given property. The size of this
range, then, is $s(\theta)$. (In set-theoretic terms, $s(\theta)$ is
the 1-D Lebesgue measure of the set of all values of $b$ that
correspond to the property in question, for fixed $q$, $a$, and
$\theta$.)

We can therefore compute $s(\theta)$ for any given property by
generating many different source paths with incidence angle $\theta,$
and counting the fraction of all events in which the light curve
exhibits that property.  Specifically, we generate light curves for
many values of $b$ for each value of $\theta$.  For each such
trajectory, we sample the magnification at a sequence of times $t_i$,
compute the corresponding $A_i=A(t_i)$, and examine the resulting
light curve. The sampling rate $\Delta t$ is the difference between
consecutive values of $t$. As a default we choose $0.01 \theta_E$, but
in \S \ref{robustness} we explore how the results vary for different
sampling rates.

Once we have determined $s(\theta)$, the next step is to average over
the possible values of $\theta$: $\langle s\rangle = \frac{1}{2\pi}
\int_0^{2\pi} s(\theta)\,d\theta$. For some event types, there is a
simple expression for $s(\theta)$, so we may compute $\langle
s\rangle$ analytically.  Generally, however, no such simple expression
exists, so we must perform the integration numerically with
simulations.

From the values of $\langle s\rangle$ for various properties, we may
compute the relative rates for light curves with those properties. For
instance, suppose that for a fixed $q$ and $a$, the value for
caustic crossing events is $\langle s_{cc} \rangle$, and the value for
all events (i.e.  light curves that exceed our minimum cutoff
magnification $A_{\rm cut}$) is $\langle s_e \rangle$. Then, for this
binary alone, the probability of an event being a caustic crossing
event is $\langle s_{cc} \rangle / \langle s_e \rangle$, and this also
eqquals the relative rates for caustic crossing events to all binary
events. Realistically, however, there is a population of binary lenses
described by a probability distribution function $P(q,a)$. In that
case, the relative rate we seek is given by $\int
\langle s_{cc}(q,a) \rangle P(q,a)\,dq\,da / 
\int \langle s_e(q,a) \rangle P(q,a)\,dq\,da$. Our results for individual
binaries are given in \S \ref{individual-results}, and our results for
populations of binaries are given in \S \ref{population-results}.

\subsection{Caustic crossing $\langle s\rangle$}
\label{caustic-crossing-s}

For caustic crossing light curves, there is a simple analytic
expression for $s(\theta)$ and thus for $\langle s\rangle$. This is
because caustic crossing light curves are defined by whether the
trajectory crosses the caustic region. We may therefore determine
whether a trajectory corresponds to a caustic crossing light curve
without generating the light curve itself. So for caustic crossings,
$s(\theta)$ is the width of the caustic region as seen from the angle
$\theta$ (see Figure \ref{geometry-fig}), and $\langle s\rangle$ is the
average width of this region.

It can be shown that the average width of any convex region (and thus
$\langle s\rangle$ for any property that corresponds to crossing the
region) is equal to $\frac{1}{\pi}$ times the region's perimeter. For
example, not surprisingly, for a circular region, $\langle s\rangle$
is $\frac{1}{\pi}$ times the circle's circumference, that is, its
diameter. For a concave region like a caustic region, we only need to
draw the smallest convex shape that encloses the region, and then,
since every trajectory that crosses this region will cross this convex
shape and vice versa, $\langle s\rangle$ is equal to $\frac{1}{\pi}$
times the perimeter of this enclosing shape.

Caustics are everywhere concave except at a small nuber of points
called cusps, so the bounding shape is a polygon (as shown in Figure
\ref{geometry-fig}), and $\langle s\rangle$ can be computed directly 
from the locations of the cusps.  Specifically, it is:

\begin{equation}
s(\theta) =
\frac{1}{\pi} \sum_{p=0}^{n-1}\sqrt{(x_{p+1}-x_p)^2+(y_{p+1}-y_p)^2}
\end{equation}

Here the $(x_p,y_p)$ are the locations of the cusps that are vertices
of the bounding polygon and $n$ is their count; for a single caustic
region $n$ will be either 5 or 6 (and for simplicity of notation, let
$(x_n,y_n) = (x_0,y_0)$). The cusp locations are given by the
simultaneous roots of two polynomials of degree 11 and 10 (see
\citet{SchneidersBook} Eq. 6.23). These roots can be computed to
sufficient precision with minimal difficulty in a program such as
Mathematica.

For a binary lens, there are either one, two, or three separate
caustics. The above analysis strictly applies only to binaries with a
single caustic, but it can easily be made to apply to the other cases
as well. For binaries with two caustics, we need only to compute
$\langle s\rangle$ for trajectories that cross one or the other or
both: if $s_1(\theta)$ corresponds to crossing the first caustic
region, $s_2(\theta)$ corresponds to crossing the second, and
$s_{12}(\theta)$ corresponds to crossing both, then corresponding to
the overall feature of crossing one or the other or both is $s(\theta)
= s_1(\theta) + s_2(\theta) - s_{12}(\theta)$ (because $s(\theta)$ is
the measure of the intersection of the sets for which $s_1(\theta)$
and $s_2(\theta)$ are measures).  Thus for caustic crossing events,
$\langle s\rangle = \langle s_1\rangle + \langle s_2\rangle -
\langle s_{12}\rangle$. The value $\langle s_{12}\rangle$ can also be
expressed in terms of the concave bounding shapes of the two regions,
so it may be computed analytically as well. A similar formula holds
for binaries with three caustics.

Thus we can easily compute $\langle s\rangle$ for caustic crossing
light curves to a high precision for all $q$ and $a$. While this does
lend some degree of insight, we use it primarily to check the
simulations that compute it and other values numerically.

\subsection{The value of $\langle s\rangle$ for other light curve features}
\label{general-case-of-s}

There are other type of events that lend themselves to analysis almost
as simple as the analysis for caustic crossings. For instance, if we
say a light curve contains an event when its maximum magnification is
greater than a threshold value $A_{\rm cut}$, then $\langle s\rangle$ for
events is $\frac{1}{\pi}$ times the perimeter of the smallest convex
shape that bounds the isomagnification contour for $A_{\rm cut}$. For wide
binaries, repeating events (those which dip below $A_{\rm cut}$ and back
up) are another example (Di\thinspace Stefano \& Mao 1995).  When the
isomagnification contour for $A_{\rm cut}$ comprises two separate convex
curves, as is the case with well separated binaries, $\langle
s\rangle$ for repeating events can be determined as was done for
disconnected caustics above, simply from the coordinates of the points
in the isomagnification contour corresponding to $A_{\rm cut}$. For
multipeaked events (those with two or more local maxima), even though
there is not a simple expression for $\langle s\rangle$ in terms of
the perimeter of a region, a sense of whether a given trajectory will
generate a multipeaked light curve can be obtained by examining how it
crosses the isomagnification contours.

In general, however, there is no easier way to compute $\langle
s\rangle$ than numerically.  For example, Figure \ref{btheta} shows
that $s(\theta)$ for multipeaked events and asymmetric events are more
complicated than for caustic crossing events.

\section{Simulation Specifications}
\label{simulation-specifications}

Our simulation generates a set of light curves for a given $q$ and $a$
by selecting values of $b$ and $\theta$, one at a time. The light
curve is defined by a discrete set of times $t_i$ and the
corresponding amplifications $A_i = A(t_i)$. If the light curve
contains an event (${\rm max}(A_i) > A_{\rm cut}$), certain light
curve parameters are measured. Our default value of $A_{\rm cut}$ is
1.1, that is, a 10\% magnification above baseline. (For comparison
with such surveys as OGLE, $A_{\rm cut} = 1.34$ may be more
appropriate; see \S \ref{robustness}.) Among light curves with events,
running totals of all light curves with certain parameter values and
characteristics are kept.

\subsection{Trajectory selection}
\label{trajectory-selection}

\begin{figure}[ht]
\begin{center}
\includegraphics[width=3in]{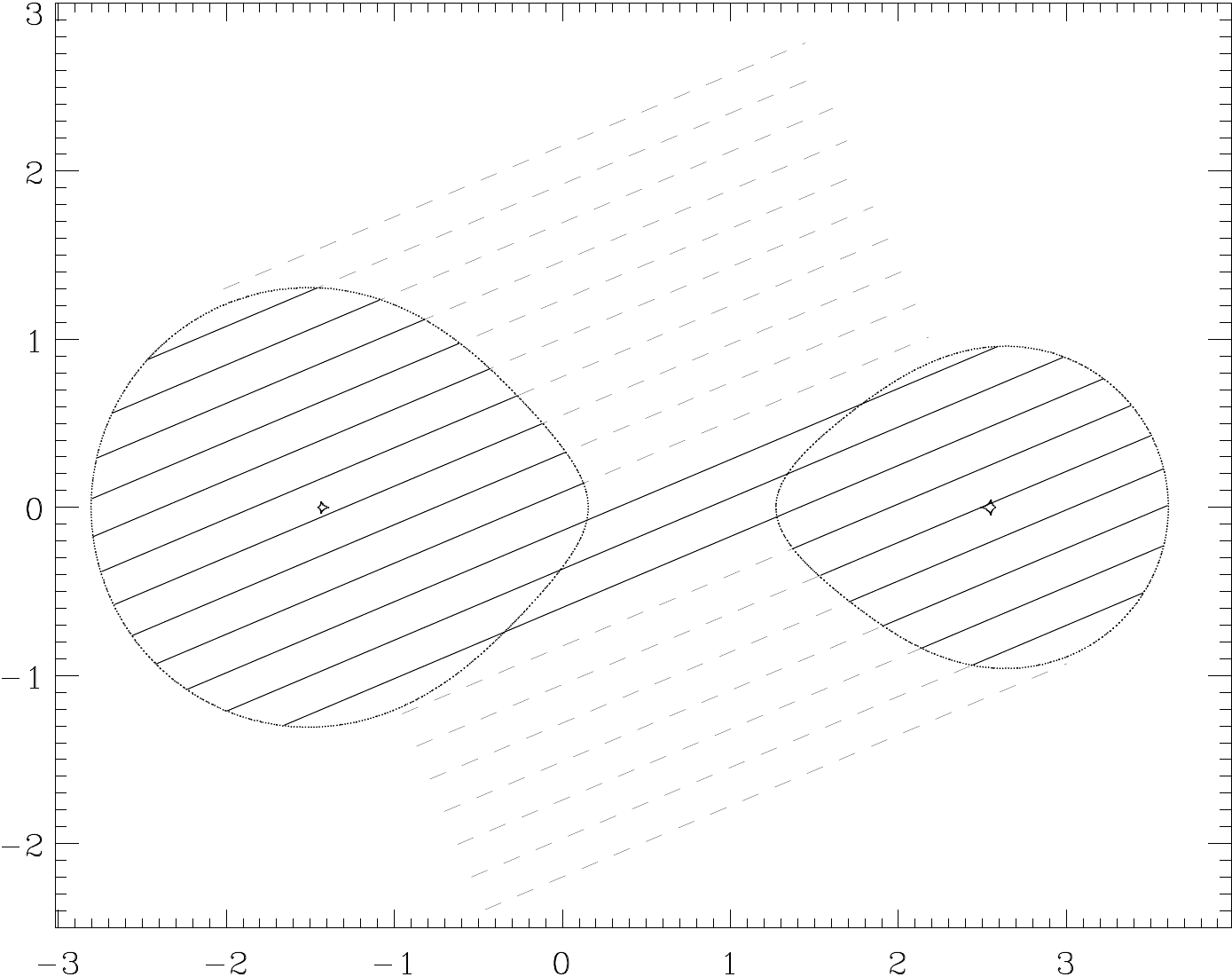}
\end{center}
\caption{
Coverage of sampling. The $x$-$y$ plane shown is the lens plane,
measured in units of $\theta_E$, as in Figure \ref{geometry-fig}. For
$(q,a,\theta) = (0.562, 4.217, 0.400)$ (chosen as a typical example),
each dashed line is a trajectory for a light curve that was generated,
each with a different value of $b$. Solid lines correspond to the
parts of the trajectory that were identified as events.
Isomagnification contours corresponding to $A_{\rm cut} = 1.1$ and
locations of the caustics are also shown. The sampling in $b$ used in
the simulations is 50-100 times denser than that shown in this figure.
}
\label{coverage}
\end{figure}

We wish to select values of $b$ and $\theta$ in such a way as to mimic
the way they are randomly ``selected'' in nature. This is done simply
enough by realizing that the probability distribution function for the
trajectories is uniform in both $b$ and $\theta$.
We select 1600
equally spaced values for $\theta$ between 0 and $\pi/2$, and for each
of these determine the set of values for $b$ that will completely
cover the set of events. This set is all trajectories that cross the
isomagnification contours of $A(x_s,y_s)=A_{\rm cut}$ in the lens
plane, as shown in Figure \ref{coverage}.

Although it is essential to completely cover these isomagnification
contours, it is desirable to limit the number of light curves
generated that do not cross the contours.  Starting with a given
binary, defined by the values $(q,a)$, and considering an individual
value of $\theta,$ there is a limited range of values of $b$ that
yield events.  For a given $\theta,$ we first identify a value of $b$
for which we know there is an event.  We accomplish this by starting
with the fact that, for any value of $A_{\rm cut},$ there are
isomagnification contours that enclose caustics, as in Figure
\ref{coverage} with two separate contours, each containing one caustic. 
We therefore begin sampling $b$ by choosing a value that hits a
caustic cusp. We then continue by choosing a sequence of values of
$b$, separated by $\Delta b$, from the original one. We sample in both
perpendicular directions (``up'' and ``down'') from the original
trajectory, keeping $\theta$ fixed.  Therefore, by starting at each
caustic and moving in both directions by an amount $\Delta b$ (equal
to 0.002 for most binaries in the simulation) for as long as events
are generated, we can be sure to cover the entire contour, while
limiting non-event light curves. (This corresponds to stopping when
the white area is reached in the left panel of Figure \ref{btheta}.)
The caustics, as mentioned in the previous section, are bounded by
their cusps, which can be located to high precision before the
simulation is run. Double sampling is avoided by maintaining a running
list of the range of $b$ sampled so far.

For each event, as the source gets far enough away from the lens, the
magnification drops toward 1, so for any trajectory there is a certain
range of $t$ for which $A(t) > A_{\rm cut}$. The limits of $t$ should
be chosen to at least cover this range, but because the tails with
$A(t) < A_{\rm cut}$ are ignored, it is a good idea to reduce these
limits as much as possible while still ensuring that all regions above
$A_{\rm cut}$ are covered. Therefore, as with $b$, we chose to stop
light curve generation when $A(t) < A_{\rm cut}$ {\em and} all the
caustics have been passed (notice the rectangular shape of the dashed
lines in Figure \ref{coverage}). Starting from $t = 0$ and moving in
both directions, this allows us to avoid computing many unneeded
magnification values.

\subsection{The magnification function}
\label{magnification-function}

The time consuming part of light curve generation is determining the
value of the magnitude function $A(q,a,x_s,y_s)$. To compute it, one
determines the locations in the lens plane of the lensing images, from
which $A$ can be computed analytically. The positions $(x_i,y_i)$ of
the images, however, cannot be computed analytically. They are
expressed in terms of the simulataneous roots of two polynomials in
$x_i$ and $y_i$ of degree 5 and 4, or alternately in terms of a single
complex polynomial of degree 5 \citep{Erdl1993}.

The method we choose to determine these roots numerically is Newton's
Method with polynomial deflation \citep{NumericalRecipes} on the real
fifth-degree polynomial given in \citep{Asada2004}.\footnote{The relevant
equations in Asada et al are 2.25-2.34. Note that what are here called
$(m, a, x_s, y_s)$ are there called $(\nu, l, a, b)$. Note the
following two typos: Eq. 2.32 is missing a factor of $b$ outside the
parentheses, and Eq. 2.33 is missing an overall factor of $b^2$.}
Because it uses real (rather than complex) arithmetic, this method,
combined with some other minor optimizations, allows for a great
improvement in speed over root finding for the complex polynomial.

Finally, after up to several hundred calls to the function, the light
curve is complete, and stored as an array of $t_i$ and $A_i$
values. We may then analyze it, and record its characteristics and
statistical parameters.

\subsection{Defined parameters}
\label{defined-parameters}

We seek to quantify the amount by which a binary lens light curve
differs from a point lens light curve. To this end we define three
parameters. For a point lens light curve, the value of each parameter
is exactly zero, and light curves with more pronounced deviations from
the point lens form have higher parameter values. Our three parameters
relate to the least-squares fit of a point lens, the asymmetry of the
curve, and whether the light curve exhibits more than one local
maximum, i.e., whether it is multi-peaked.

\subsubsection{Best fit $\sigma_P$}
\label{best-fit-sigma}

The most straightforward way to evaluate whether a light curve is
consistent with a certain model is to fit it to that model using a
least-squares fit. Therefore we begin by fitting each binary lens
light curve with a standard 3-parameter point lens form, known as the
Paczy{\'n}ski model. (We assume that in observations, the baseline can be
determined arbitrarily well, so it is not considered a fit parameter.)
As usual, we define the closest point lens model, $A_{\rm PL}(t)$, to be
the one which minimizes the sum of the squared differences between the
values of the simulated binary lens light curve, $A_i,$ and the model.

\begin{equation}
\sum_{i=1}^n\Big(A_i -A_{\rm PL}(t_i)\Big)^2
\end{equation}

The binary lens light curve offers a certain amount of intrinsic
deviation from the point lens model, characterized by this
sum-of-squares quantity. In real observations, there will be
additional deviation due to the photometric uncertainty. Roughly
speaking, when the intrinsic deviation is larger than the photometric
deviation, the binary nature of the light curve will be evident
through least-squares fitting. When the intrinsic deviation is smaller,
$A_{\rm PL}$ will provide an acceptable fit to the binary lens light curve.

Thus we define our first parameter, $\sigma_P$, as a measure of how
large the photometric uncertainty would need to be in order for the
observed light curve to be well fit by a point lens model. This
parameter is given by the root-mean-squared difference between the
simulation light curve and the best fit light curve:

\begin{equation}
\sigma_P = \Big[\frac{1}{n-3}\sum_{i=1}^n \Big(A_i - A_{\rm PL}(t_i)\Big)^2\Big]^{1/2}
\end{equation}

Note that $\sigma_P$ does not actually quantify the goodness of fit of
a point lens model to a simulated binary lens light curve; since there
are no uncertainties associated with the simulated light curve, such a
statistic is meaningless.  For an observed light curve with associated
uncertainties, the goodness of fit is quantified by the reduced
$\chi^2$ test statistic.  Its formula is very similar to our formula
for $\sigma_P$:

\begin{equation}
\chi^2 = \frac{1}{n-3}\sum_{i=1}^n \frac{\big(A_i - A_{\rm PL}(t_i)\big)^2}{\sigma_i^2}
\end{equation}

The $t_i$, $A_i$, and $\sigma_i$ are the times, magnifications, and
uncertainties of the observed light curve, of which there are $n$.
$A_{\rm PL}(t)$ is the three parameter point lens fit chosen such that
$\chi^2$ is minimized. For our purposes, we will assume that the
uncertainties $\sigma_i$ are equal in magnitude space to a general
survey parameter $\sigma_{phot}$ that quantifies the photometric
uncertainty of the survey.

\begin{figure*}[thbp]
\begin{center}
\includegraphics[width=5in]{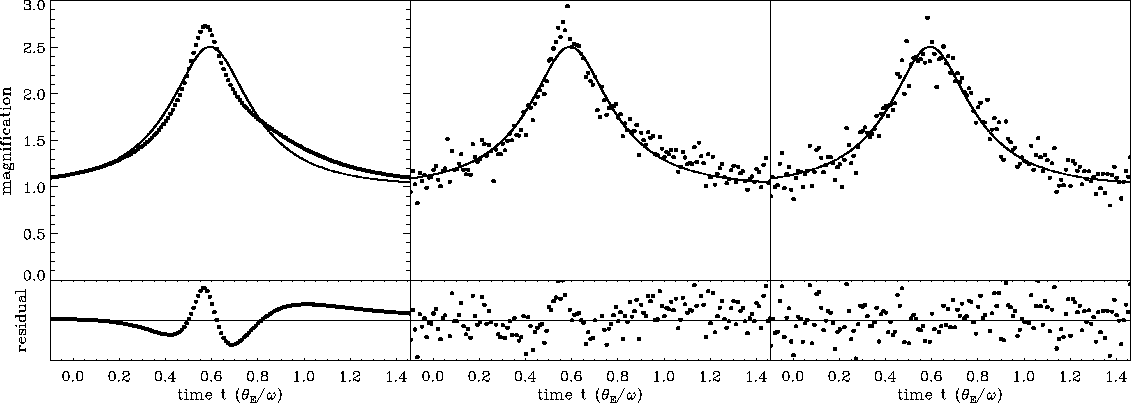}
\end{center}
\caption{
Comparison of residuals for an asymmetric binary lens light curve and
a point lens light curve. The binary lens light curve in the left
panel (dots, corresponding to simulated observations) has a best fit
point lens light curve as shown by the solid curve. It has a best fit
parameter of $\sigma_P = 0.09$, the root-mean-square deviation from
this fit, but an asymmetry parameter of $k = 0.21$. The center panel
shows the same observations with $10\%$ Gaussian error added; if this
light curve were fit assuming $10\%$ photometric uncertainty, the
best fit test statistic would be $\chi^2 < 2$. The right panel shows
the same best fit point lens, with enough Gaussian error added to give
it a comparable $\chi^2$.  Below each light curve are shown the
residuals after subtracting out the best fit curve. Although the
average size of the residuals is comparable in the two right panels,
the residuals of the center panel show much more apparent structure,
as expected.  }
\label{asym-compare}
\end{figure*}

\begin{figure*}[th]
\begin{center}
\includegraphics[width=5in]{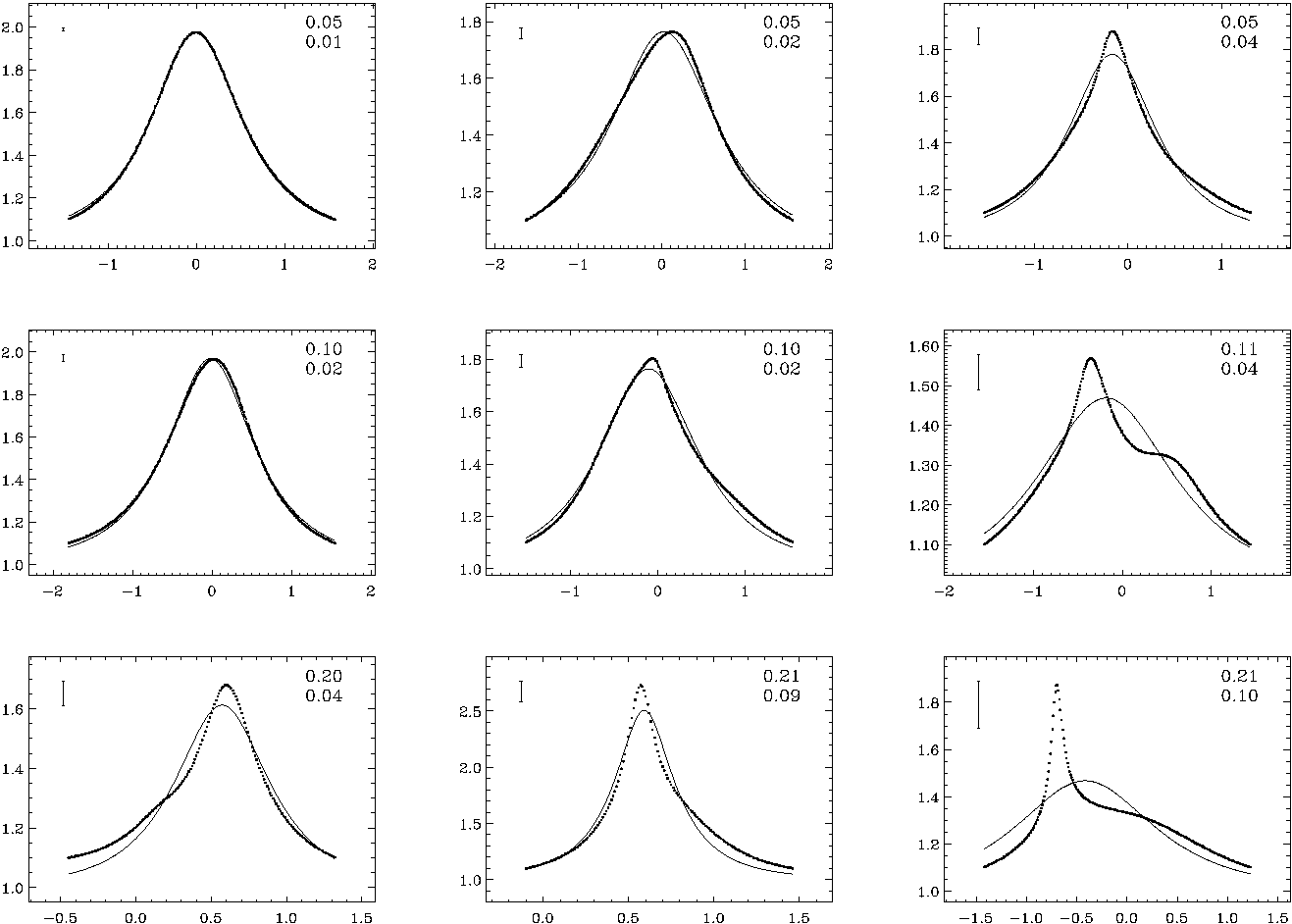}
\end{center}
\caption{
A variety of binary lens light curves along with their best fit
point lens models. In the upper right of each panel is shown the
asymmetry parameter $k$ followed by the best fit parameter
$\sigma_P$. The top, middle, and bottom rows show light curves with
asymmetry parameter $k$ equal to approximately 5\%, 10\%, and 20\%.
The error bar in the upper left has a value of $2\sigma_P$; thus it is
the size of the error bar necessary to make the best fit $\chi^2$
equal unity. Generally speaking, the greater the parameter values, the
further the light curves are from their point lens models.  }
\label{curves}
\end{figure*}

Thus, roughly speaking, for a given binary lens light curve,
$\sigma_P$ is the critical value for the photometric uncertainty: if
$\sigma_{phot} > \sigma_P$, the light curve will be well fit by a
point lens model, and if $\sigma_{phot} < \sigma_P$, it will not.
Strictly speaking, if a binary lens light curve with intrinsic
deviation $\sigma_P$ is observed with uniform, uncorrelated, Gaussian
random photometric errors with standard deviation $\sigma_{phot}$, and
then fit with a point lens model, the reduced $\chi^2$ statistic will
have an expected value (for $n \gg 1$) of $\chi^2 = 1 +
(\sigma_P/\sigma_{phot})^2$.  Therefore, for $\sigma_P >
\sigma_{phot}$, the expected value of $\chi^2$ exceeds 2, so the
point lens model would probably be identified as a bad fit. This is
the basis of our rule of thumb that a survey with an average photometric
uncertainty of $\sigma_{phot}$ should discriminate as non-point lens a
binary lens light curve with $\sigma_P > \sigma_{phot}$.

\subsubsection{Asymmetry parameter $k$}
\label{asymmetry-parameter}

Model fitting reveals how much a binary lensing light curve differs
from the point lens model, but not in what way. Because point lens
light curves are always symmetric in time, i.e., $A(t)$ is an even
function, one obvious way in which we expect some binary lens light
curves to differ is by exhibiting some asymmetry with respect to time
reversal.  Therefore we define an asymmetry parameter $k$ that is 0
for a symmetric light curve, and larger for more asymmetric ones. The
parameter $k$ is based on the Chebyshev coefficients $T_n$, which are
the convolution of the light curve function $A(t)$ with Chebyshev
polynomials, as in \citet{DiStefanoPerna}. For even $n$ the
polynomial is an even function, and for odd $n$ the polynomial is an
odd function. Therefore a function with even symmetry has $T_n = 0$
for all odd $n$, and a function with some asymmetry must have $T_n$
significantly nonzero for some odd $n$. We then define the asymmetry
parameter to be the ratio of the root-mean-square average of the odd
coefficients to the even coefficients:

\begin{equation}
k = \Big({\displaystyle\sum_{n=1}^\infty T_{2n+1}^2}\Big/{\displaystyle\sum_{n=1}^\infty T_{2n}^2}\Big)^{1/2}
\end{equation}

One use of the asymmetry parameter is to characterize the way in which
a particular light curve deviates from the point lens form.  We note,
however, that when we apply model fits to light curves with pronounced
asymmetries, the residuals to the best point lens fit are not randomly
distributed in time, but instead exhibit some structure. That is, the
deviations from the point lens light curve are correlated, even when
$\sigma_P$ is small.  Thus the parameter $k$ can identify light curves
with small but correlated residuals, whereas the parameter $\sigma_P$
cannot.  An example of two light curves with identical $\chi^2$ but
different $k$ is shown in Figure \ref{asym-compare}.

In this paper we identify light curves that are smoothly perturbed
from a point lens form by their values of $\sigma_P.$ We note,
however, that in future work it may be possible to use the asymmetry
parameter to identify binary lens light curves, even if the
best fit parameter fails to distinguish them from point lens events.
Figure \ref{curves} shows typical light curves with various
intermediate values of $\sigma_P$ and $k$.  Extremely point lens-like
or extremely perturbed events will easily stand out with any choice of
statistics.

\subsubsection{Multipeak parameter $p$}
\label{multipeak-parameter}

This parameter is identically 0 for any light curve with a single
peak. For light curves with two peaks, it is the difference between
the first peak and the intervening minimum, or the difference between
the second peak and the intervening minimum, whichever is smaller. The
idea is that if differences of less than $p$ were undetectable, the
multipeaked nature of the light curve would go unnoticed. For light
curves with three or more peaks, it is the largest such value for any
of the minima in the light curve:
\begin{equation}
p = {\rm Max}_i({\rm Min}(A_{{\rm max},i},A_{{\rm max},i+1})-A_{{\rm min},i})
\end{equation}
where $A_{{\rm min},i}$ is the magnitude at the $i$th minimum and $A_{{\rm
max},i}$ is the magnitude at the $i$th maximum.

\section{Results}
\label{results}

\subsection{Results for individual binaries}
\label{individual-results}

\begin{figure*}[thp]
\begin{center}
\includegraphics[width=5in]{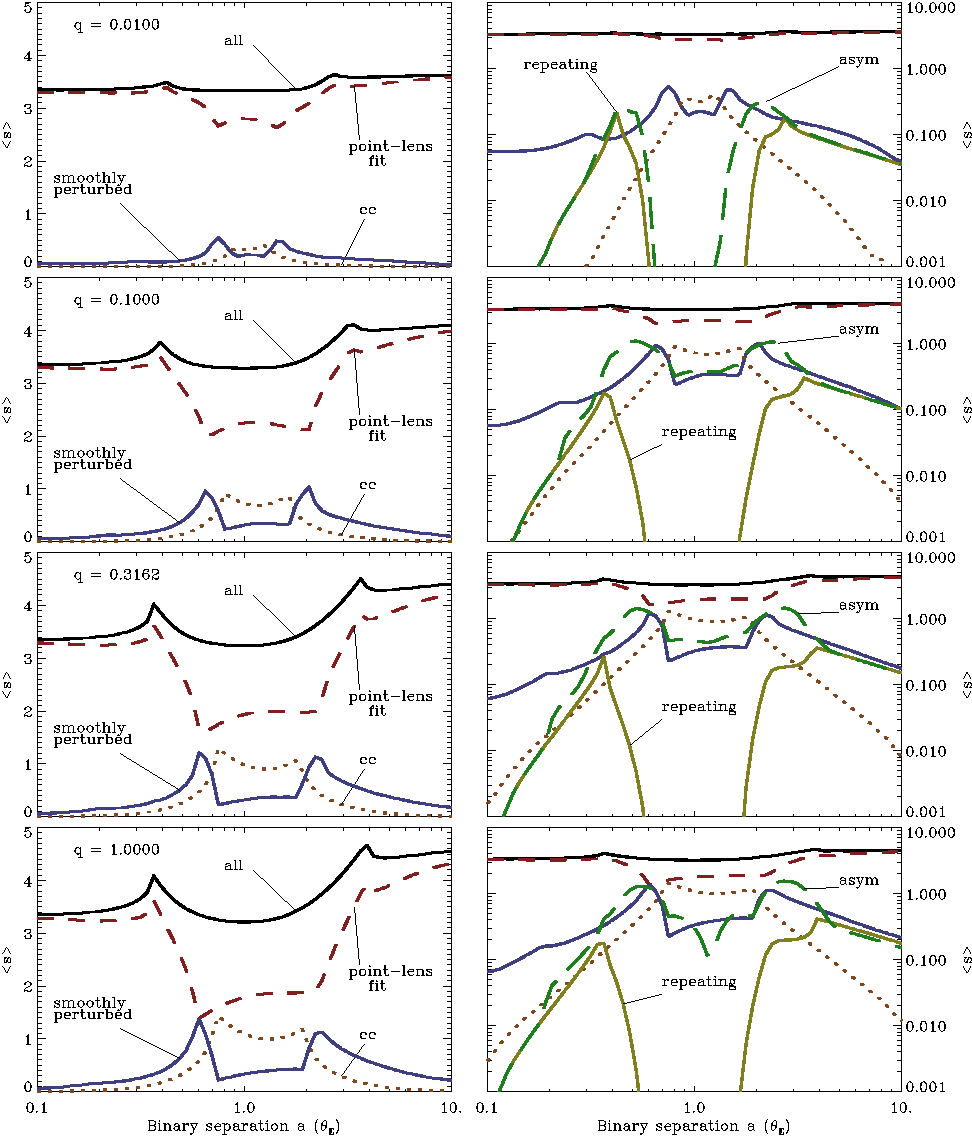}
\end{center}
\caption{
The relative rate measure $\langle s\rangle$ for various types of
binary lensing events for certain fixed values of $q$. For each of the
four values of $q$ as indicated in the left panels, $\langle s\rangle$
is plotted against $a$. The types of events identified are: caustic
crossing events (orange dotted curve), smoothly perturbed events
($\sigma_P > 0.1$, blue solid curve), point lens like events
($\sigma_P < 0.1$, red dashed curve), and the total of these three
categories (black solid curve). Note particularly that there are many
values of $q$ and $a$ where the smoothly perturbed events outnumber
the caustic crossing events. The right panels show the same data as
the left panels, except on a logarithmic plot. The right panels also
show the relative rates for two more types of events, repeating events
($A_{\rm min} < A_{\rm cut}$, solid yellow curve) and asymmetric
events ($k > 0.1$, dashed green curve). There is some overlap between
these two categories, and between these categories and the others.  }
\label{profile}
\end{figure*}

\begin{figure*}[ht]
\begin{center}
$ \begin{array}{cc}
\includegraphics[width=3in]{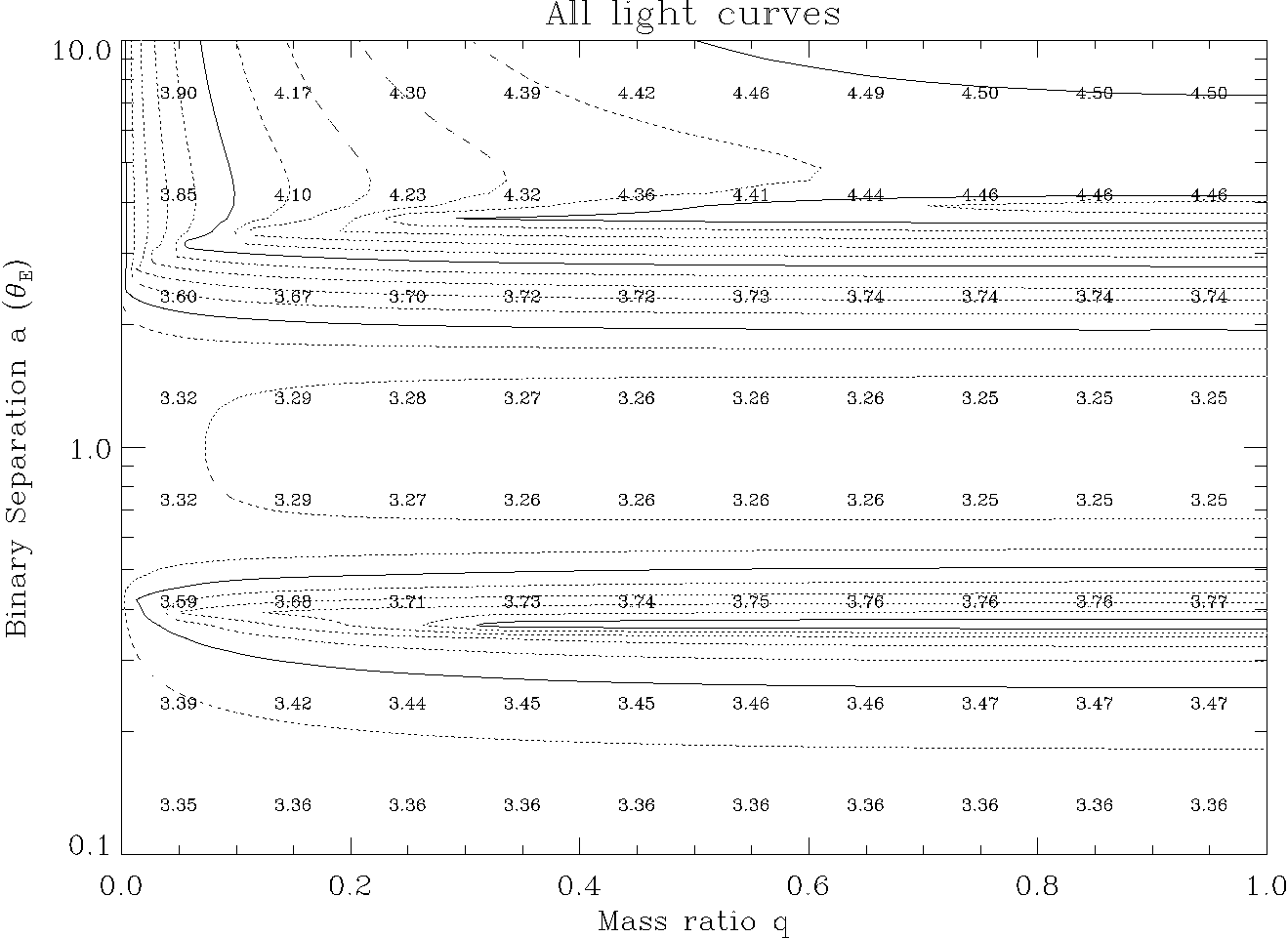} &
\includegraphics[width=3in]{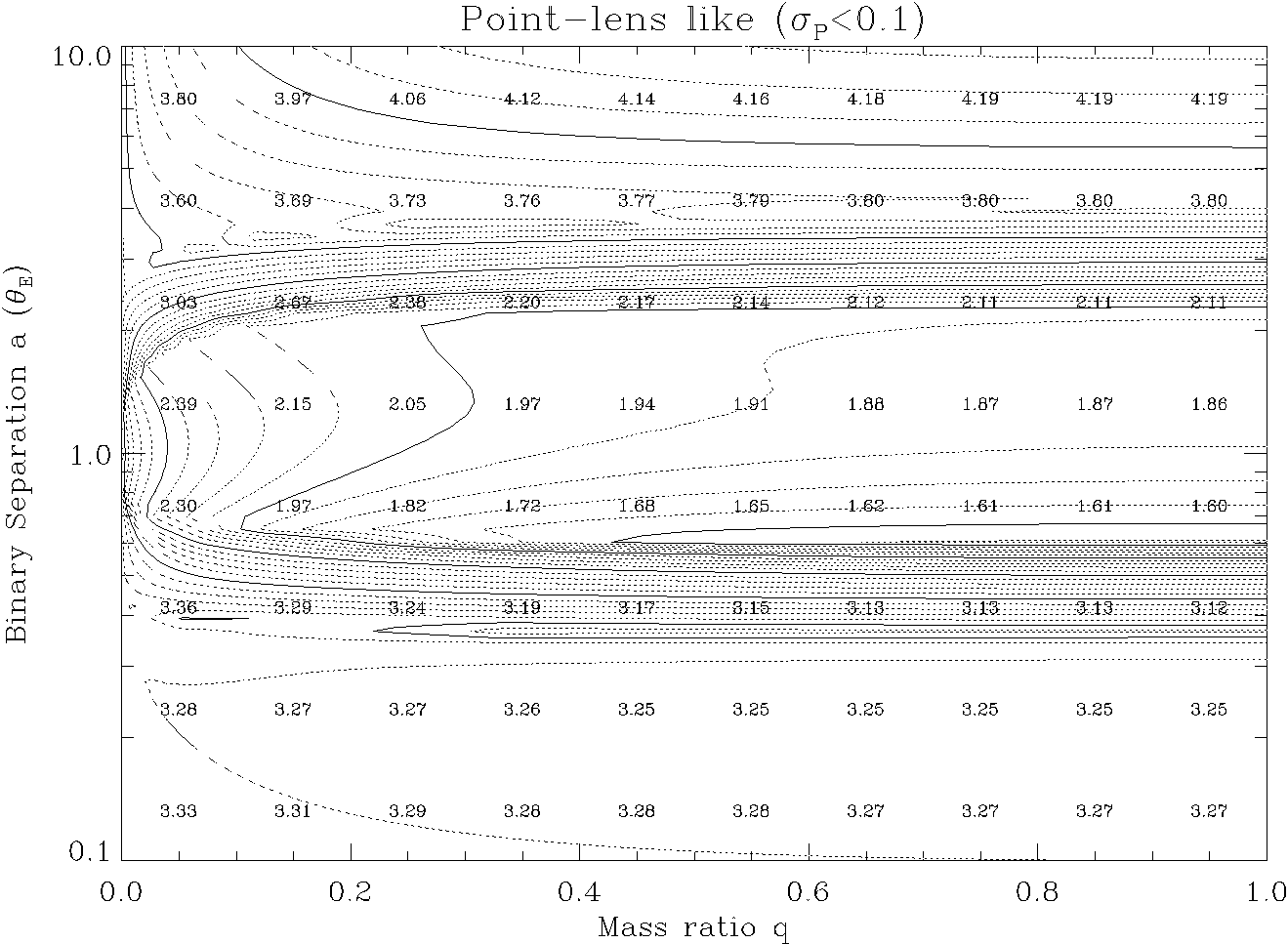} \\
\includegraphics[width=3in]{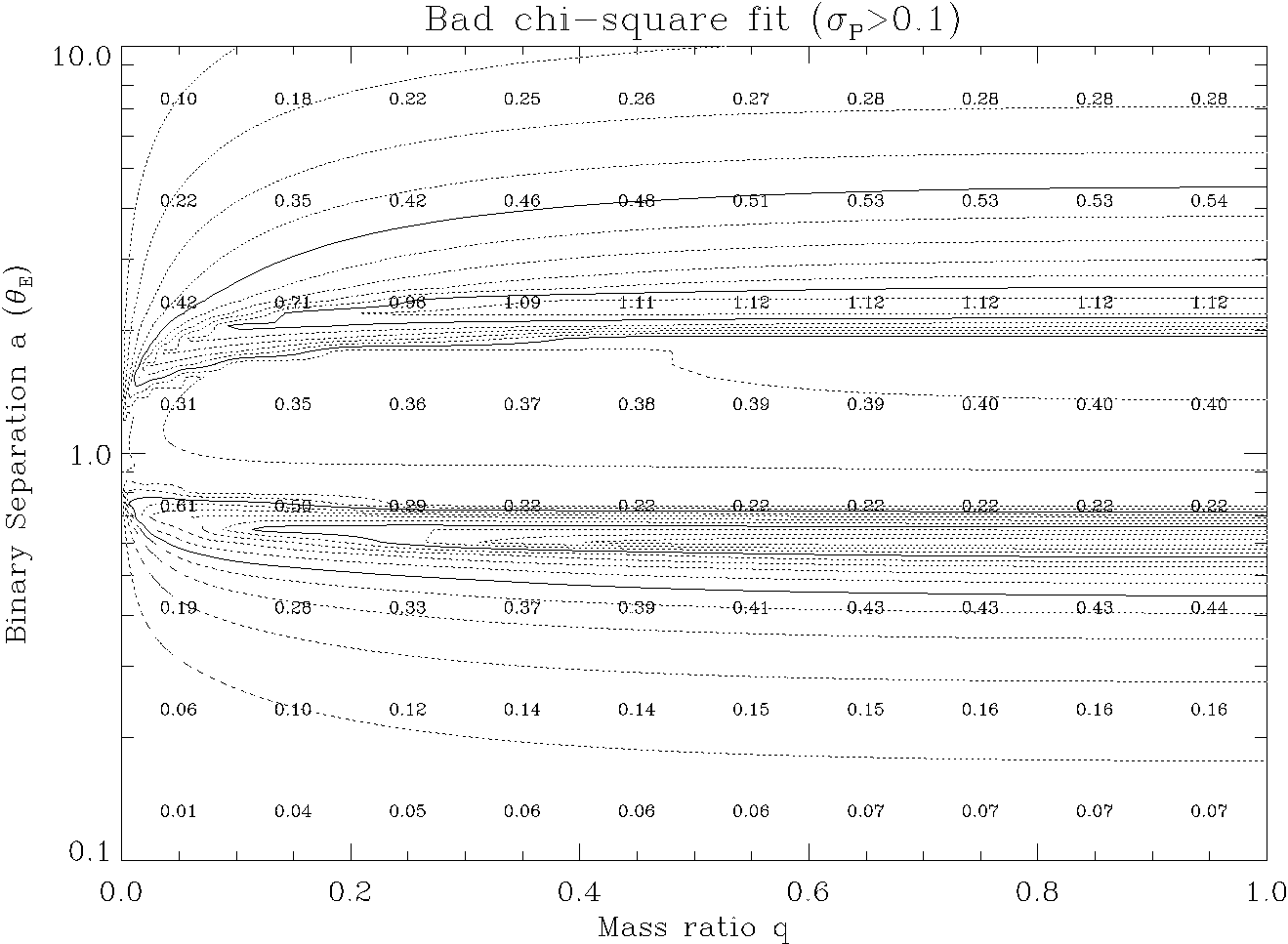} &
\includegraphics[width=3in]{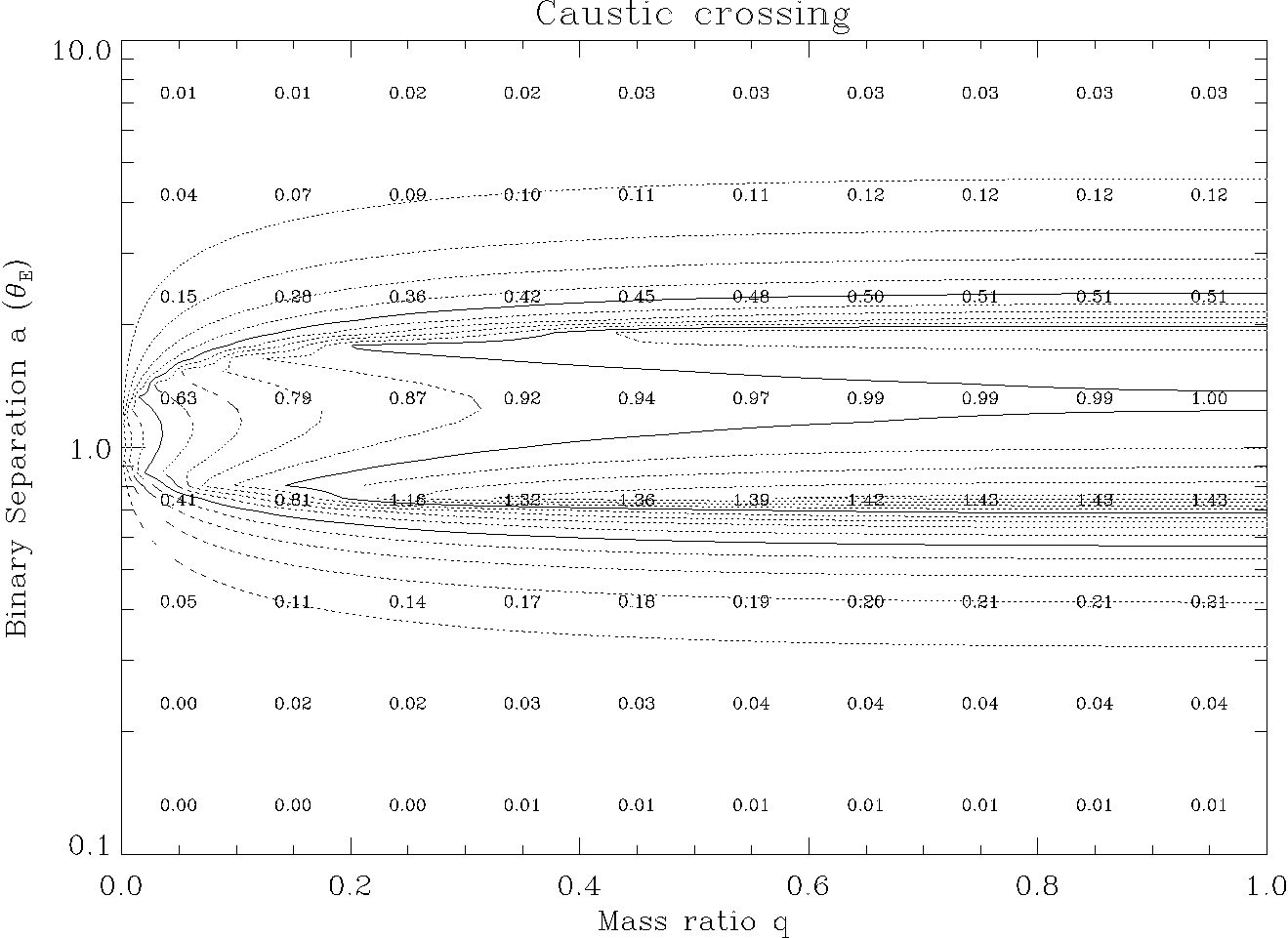} \\
\includegraphics[width=3in]{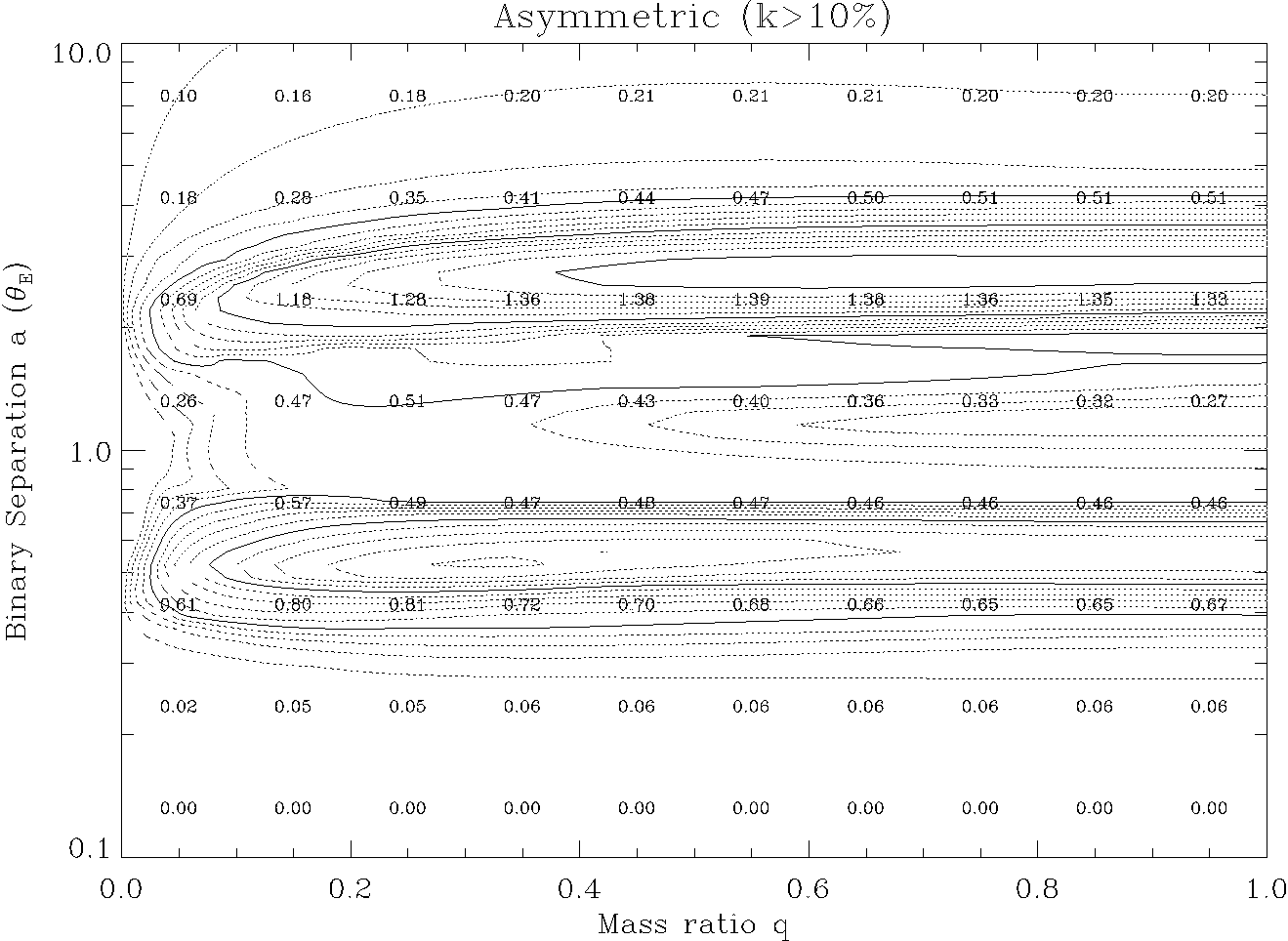} &
\includegraphics[width=3in]{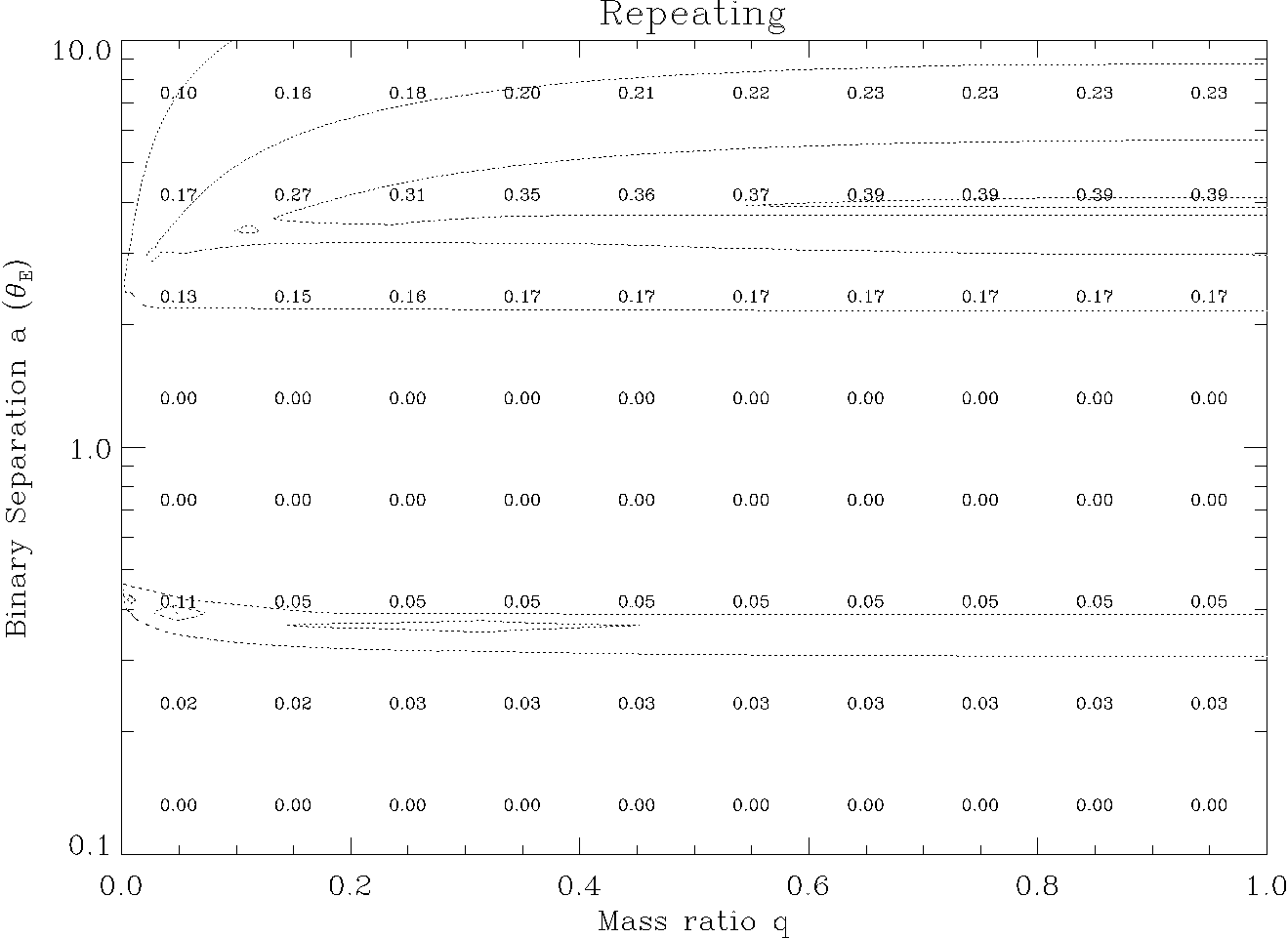}
\end{array} $
\end{center}
\caption{
Contour plots of $\langle s\rangle$ vs. $q$ and $a$ for the entire
range covered by our simulations. Each panel corresponds to a
different type of event: all events ($A_{\rm max} > A_{\rm cut}$),
point lens like events ($\sigma_P < 10\%$), smoothly perturbed events
($\sigma_P > 0.1$), multi-peaked events at 10\% (multipeak $k > 0.1$),
and finally repeating events ($A_{\rm min} < A_{\rm cut}$).
Consecutive contours are separated by 0.1 with dark contours every
0.5. Overwritten numbers show the same data in tabular form. Notice
that generally there are more variations with varying $a$ than with
varying $q$, and that the curves exhibiting binary signatures exist
most in the ``critical zone'' around $a = 1$. For a point lens light
curve, $\langle s
\rangle$ for the total number of curves would be 3.35.}
\label{contour}
\end{figure*}

Figures \ref{profile} and \ref{contour} depict the relative event
rates $\langle s\rangle$ as a function of $q$ and $a$. Figure
\ref{profile} plots these rates for several fixed values of $q$, and Figure
\ref{contour} shows contour plots for the entire range of $q$ and $a$
covered by our simulation. The graphs in Figure \ref{profile} are
vertical cross sections of the contour plots in Figure
\ref{contour}.

The total rate of events (light curves for which $A_{\rm max}>A_{\rm cut}$) is
shown as the top, solid black curve (``all") in each panel of Figure
\ref{profile}, and in the top left panel of Figure \ref{contour}.
The values we have computed can be checked analytically both for low
values of $a,$ where the lens essentially acts as a single point lens,
and for high values of $a,$ where the lens acts approximately as two
independent point lenses. In the limit of small $a,$ the event rate is
determined simply by our cutoff magnitude. For $A_{\rm cut} = 1.34$,
corresponding to $b \le \theta_E$ (or $b \le 1$ in our units) for a
point lens, the rate would be $\langle s\rangle = 2$. For our default
value of $A_{\rm cut} = 1.1$, this rate is $\langle s\rangle = 3.35$. On
the other hand, for $a$ very large, the masses act as two separate
point lenses, each with its own Einstein radius proportional to the
square root of its mass. Thus the event rate increases by a factor of
$\frac{1+\sqrt{q}}{\sqrt{1+q}}$. This factor is maximized with a value
of $1.414$ for $q=1$, producing an overall event rate of $\langle
s\rangle =4.73$.  This is the value approached, for large $a$, by the
solid black curves in Figure \ref{profile}.

The increase in event rate for large $a$ causes an interesting effect.
Although the lensing optical depth depends only on the total mass of
the lens and not on $a$, the event rate increases from the point lens
value as $a$ increases. Thus a lens population consisting of equal
total masses of binaries and single stars would produce more
binary lens than single lens events, even though binaries and point
lenses would contribute the same amount to the lensing optical
depth. Note that most of the binary events would be indistinguishable
from point lens events.

Also plotted are rates for the two main types of binary perturbations:
non-caustic crossing but with $\sigma_P \ge 10\%$ (smoothly perturbed
events), and caustic crossing. These are shown in Figure
\ref{profile} with a dotted orange curve and a solid blue curve,
respectively, and in the middle two panels of Figure
\ref{contour}. All events that are not perturbed in either of these
two ways are shown as ``point lens like'' events, with a red dashed
curve in Figure \ref{profile}, and in the upper right panel of Figure
\ref{contour}. The rate for ``all'' events, that is, the total rate,
is the sum of the rates for smoothly perturbed, caustic crossing, and
point lens like events.

For all values of $q$, there are values of $a$ for which the smoothly
perturbed events outnumber the caustic crossing events. In Figure
\ref{profile}, this is where the blue solid curve is above the orange
dotted curve. In Figure \ref{contour}, this can be seen by comparing
the overlaid numbers at the top and bottom of the middle two
panels. An increase or decrease in $a$ from $a=1$ results in the
smoothly perturbed event rate exceeding the caustic crossing rate. The
fact that it occurs on both sides of $a=1$ suggests that a binary
distribution function weighted toward either high or low $a$ will not
tend to favor caustic crossing events over smoothly perturbed events.

Binary perturbations are most frequent close to $a=1$ for all values
of $q$, and there tends to be more variation of event rates over the
range of $a$ covered by our simulation than over the range of
$q$. This can be seen by the largely horizontal contours in Figure
\ref{contour}. 

Small $q$ corresponds to a low mass secondary, approaching the planet
in the limit $q \ll 1$. For the most part, for small $q$, both event
characteristics and event rates approach those of point
lenses. Nevertheless, there are two regions of parameter space where
planetary systems can provide distinctive lensing signatures. The
first is for values of $a$ close to unity. In this ``resonant'' regime
\citep{MaoPaczynski, GouldLoeb}, the most obvious evidence of
the existence of the planet is provided by caustic crossings, which
punctuate an otherwise point lens like light curve.  The second is for
larger values of $a$, where asymmetry in otherwise point lens like
events, and also repeating events provide evidence of the low mass
companion \citep{DiStefanoMao, DiStefanoScalzo1, DiStefanoScalzo2}. In
this paper we focus on the regime of binary stars. As $q$ increases
from near $0$, binary signatures become more frequent and more
pronounced, with relatively little effect for varying $q$ greater than
$\sim 0.2$ (see Figure \ref{contour}). Because values of $q$ near or
larger than $0.2$ are favored by binary distributions, e.g., the
distribution studied by \citet{DuquennoyMayor}, we expect the
distribution in $a$ to have a more pronounced effects than the choice
of realistic distribution in $q$.

Although the strongest binary signatures occur within an order
of magnitude of $a = 1$, there is a dip in binary signatures in the
center of this range, very near to $a = 1$. This can be seen in Figure
\ref{profile} as a dip in total event rate and a smaller dip in caustic
crossing event rate. This can be explained in terms of the geometric
considerations of the lens plane from \S \ref{lensing-geometry}. When
$a$ is very close to 1, the caustics (and thus the region of binary
perturbations) are roughly equal in size in both dimensions. As $a$
increases or decreases, the caustics stretch in one dimension or the
other, and so they have a larger extent in the lens plane. The caustic
crossing event rate peaks just as the caustics transition from
connected to disconnected, and the perturbed event count peaks farther
from $a=1$, with the value of $a$ at the peak determined by the value
of $A_{\rm cut}$.

Although our coverage in $a$ is not infinite, spanning two orders of
magnitude from $a = 0.1\theta_E$ to $a = 10\theta_E$, it includes the
range with significant binary signatures. For any value of $q$, all
binary effects will go to 0 as $a$ becomes very large or very small. In
Figure \ref{profile}, this is seen in all panels, as all curves but the
top two go to 0 in both directions. In Figure \ref{contour}, this is seen
in the lower four panels, as all tabulated values go to 0 at the top and
bottom of the panel.

Repeating events (multipeaked events with $A_{\rm min} < A_{\rm cut}$), and
asymmetric events ($k > 0.10$) are shown in the right hand (log plot)
panels of Figure \ref{profile}, as gold solid curves and dashed green
curves, respectively. Repeating events are one type of binary behavior
that does not fall off with the size of the caustics for large $a$.
Rather, two separated point lenses can still produce a significant
number of repeating events. This can be computed analytically as in \S
\ref{lensing-geometry}. If $R$ is the rate for a point lens of the same mass
($R = 3.35$ for $A_{\rm cut} = 1.1$), then to leading order in $R/a$, the
rate for repeating events is approximately $\frac{1}{2\pi} {\rm
sech}(\frac{1}{2} \ln(q)) \frac{R^2}{a}$. This expression goes as
$a^{-1}$ whereas the caustic crossing event rate goes as $a^{-2}$.

Although asymmetric and repeating events are significantly less
frequent than caustic crossing events and smoothly perturbed events,
they are expected to constitute a few percent of binary lens events.
We should therefore find them in any large sample of binary lensing
events.

\subsection{Results for populations of binaries}
\label{population-results}

\begin{figure*}[ht]
\begin{center}
\includegraphics[width=4.5in]{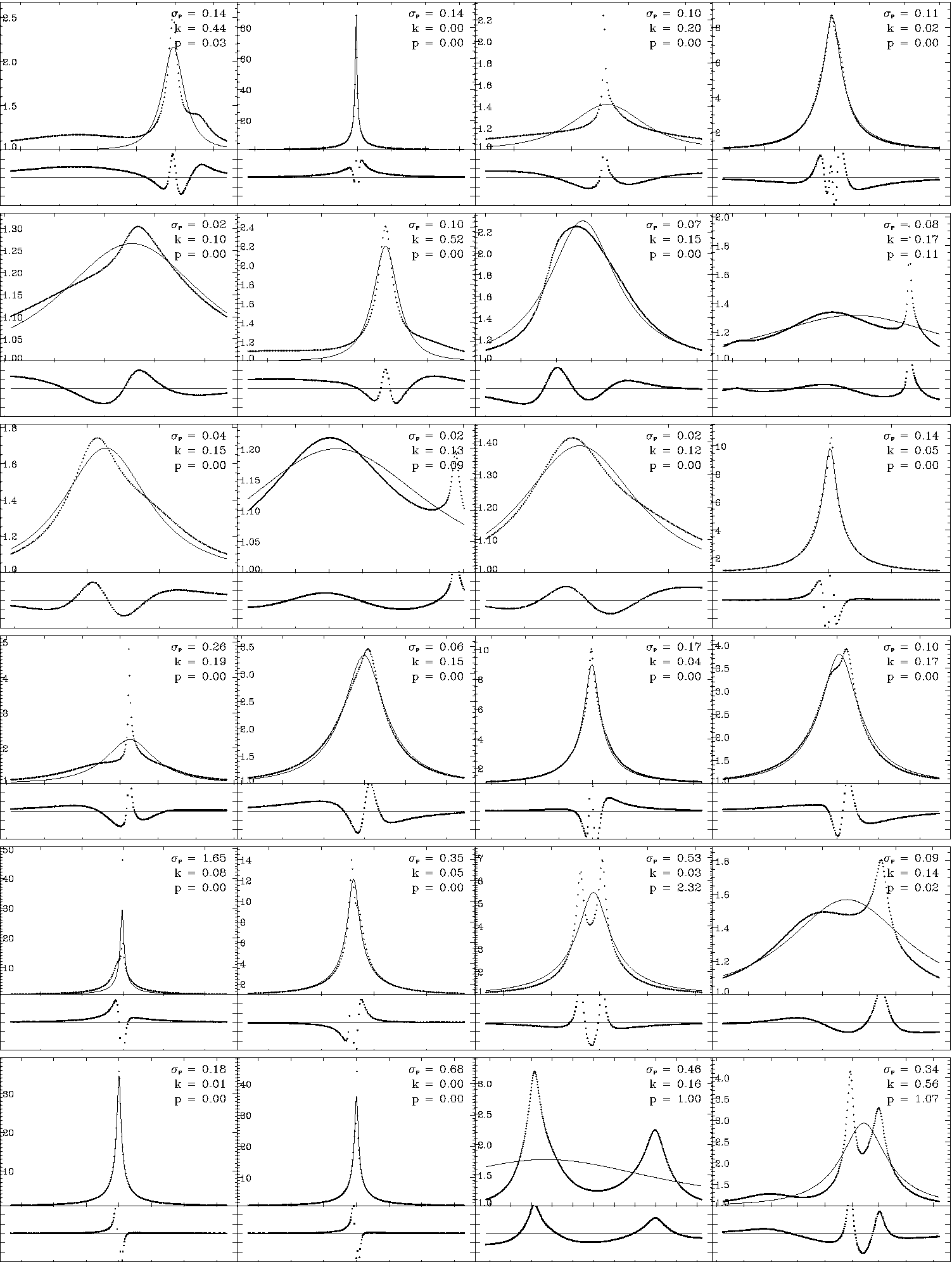}
\end{center}
\caption{
Random sampling of non-caustic crossing, nonrepeating light curves
perturbed at the 10\% level. Each light curve is shown with its
best fit point lens model. Residual plots below show the patterns of
the differences between the light curve and the model. The three light
curve parameters are shown in the upper right of each panel; these
light curves were selected from the population that has at least one
of these three parameters greater than 0.1. Horizontal (time) axis
tick marks for each light curve are in units of 0.5 times the Einstein
angle crossing time $\theta_E / \omega$. Vertical axis tick marks in
the residual plots are in units of $\sigma_P$, and the axis ranges
from $-3\sigma_P$ to $+3\sigma_P$. Our simultions show that typically,
for every 13 caustic crossing events, there are 15 events that are
smoothly perturbed with $\sigma_P > 10\%$, and a total of 24 events
for which at least one of the three parameters exceeds 0.10.
}
\label{perturb-sample}
\end{figure*}

In order to use these results to make predictions for surveys, we must
consider a population of binary lenses whose properties (mass ratios
and projected orbital separations) are drawn from a realistic
distribution given by a probability distribution function
$P(q,a)$. Then, for instance, the overall relative rate for a given
type of event is:

\begin{equation}
\langle s_x \rangle = \int\!\!\int P(q,a) \langle s_x(q,a)\rangle\,dq\,da
\end{equation}

\noindent Note that previously we have used $\langle s\rangle$ to denote
an average only over $b$ and $\theta$, but here and in subsequent
sections, we use it to denote an average over all four binary
parameters. As a default, we assume a distribution uniform in $q$ and
log-uniform in $a$, but in \S \ref{robustness} we explore how the
results vary for different distributions.

For every caustic crossing light curve, there are a certain number of
smoothly perturbed light curves, as shown in Figure
\ref{perturb-sample}. We want to find this ratio.  
Our default definition for smoothly
perturbed events is $\sigma_P \ge 0.10$, but we vary this cutoff parameter
value over a wide range, from 0.01 to 1.0, and show the ratio for all
cutoff values.

\begin{figure}[th]
\begin{center}
\includegraphics[width=3in]{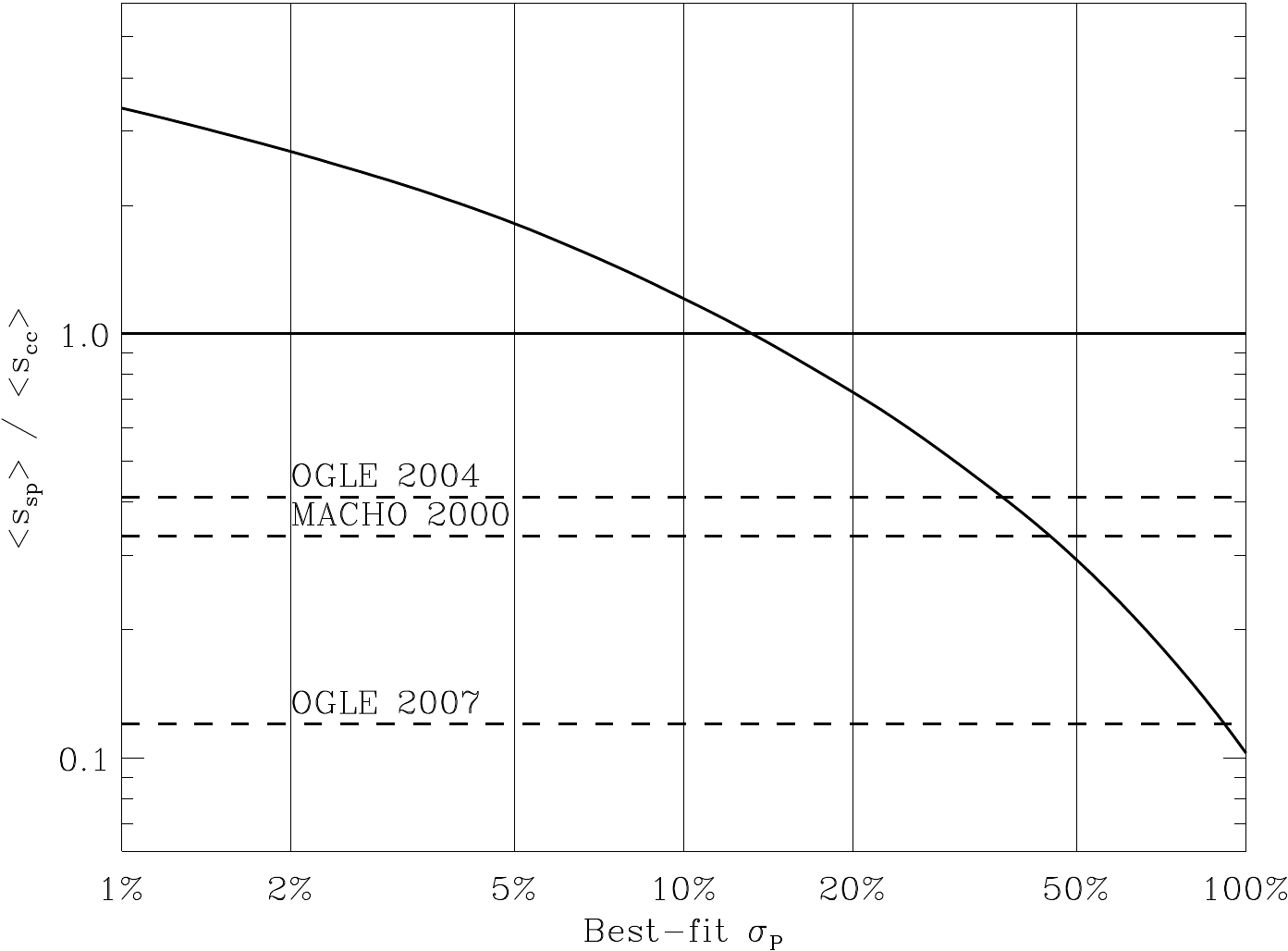}
\end{center}
\caption{ Curve showing the expected rate ratio of smoothly perturbed
to caustic crossing events. The $y$-axis shows the rate $\langle
s\rangle$ of events with $\sigma_P$ greater than the corresponding
$x$-axis value, normalized to $\langle s\rangle$ for caustic crossing
events. The horizontal solid line shows $\langle s\rangle$ for caustic
crossing events. The horizontal solid line corresponds to equal rates
of smoothly perturbed and caustic crossing events.  Under our default
assumptions, a survey with perfect detection efficiency of events
perturbed at the $\sigma_P = 15\%$ level would expect to find as many
smoothly perturbed events as caustic crossing events.  Shown as dashed
horizontal lines are the ratios reported for MACHO (5.7 year data;
\citet{MACHOBinaries}; 4 perturbed events out of 16 binaries), OGLE
III 2002-2003 season (\citet{OGLEBinaries2002}; 7 perturbed candidates
out of 24 binary candidates), and OGLE III 2004 season
(\citet{OGLEBinaries2004}; 3 perturbed events out of 25 binary
candidates). Implicit in this graph are the default assumptions for
our simulation, which are the selection criterion $A_{\rm cut} =
1.10$, the sampling rate $\Delta t = 0.01\theta_E$, the distribution
function $P(q,a)$ uniform in $q$ and $\log(a)$, and the assumption of
no blending. The following figures show how this graph is affected
when these are changed.  }
\label{sigma}
\end{figure}

Figure \ref{sigma} shows the major results. For various cutoff
$\sigma_P$ values, it shows the ratio of smoothly perturbed to caustic
crossing events. For a cutoff value of $\sigma_P$ of 0.10, this ratio
is slightly greater than unity. Roughly speaking, this means that a
survey with $10\%$ photometry should detect more smoothly perturbed
events than caustic crossing events. Naturally, the better the
photometry, the larger this ratio should be. For $1\%$ photometry, the
theoretical ratio is as high as 3. We can also answer the question:
for what precision of photometry would the observed ratios ($4/12$ for
MACHO, $7/17$ for OGLE in 2002-2003, and $3/25$ for OGLE in 2004)
match the expected ratios? In Figure \ref{sigma}, this is the value of
$\sigma_P$ where the dashed lines meet the curve, around $35\%$ or
greater.

Thus there is a large discrepancy between the values of $\langle
s_{sp} \rangle/\langle s_{cc} \rangle$ predicted by our calculations
and the values measured by present surveys, too significant to be
attributed to our use of $\sigma_P$ as an approximate measure of the
photometry needed to detect an event. We want to determine whether
this discrepancy can be explained in terms of the assumptions made in
our simulations.  While we did not assume any particular photometric
precision, we did assume default values of the sampling rate and event
detection threshold, values that are more ideal than could be expected
of a current survey. It appears, however, that the most obvious such
possible explanations cannot be responsible for the discrepancy. We
may see this by determining the effect that modifying these
assumptions has on the results.

\begin{figure}[th]
\begin{center}
$ \begin{array}{c}
\includegraphics[width=3in]{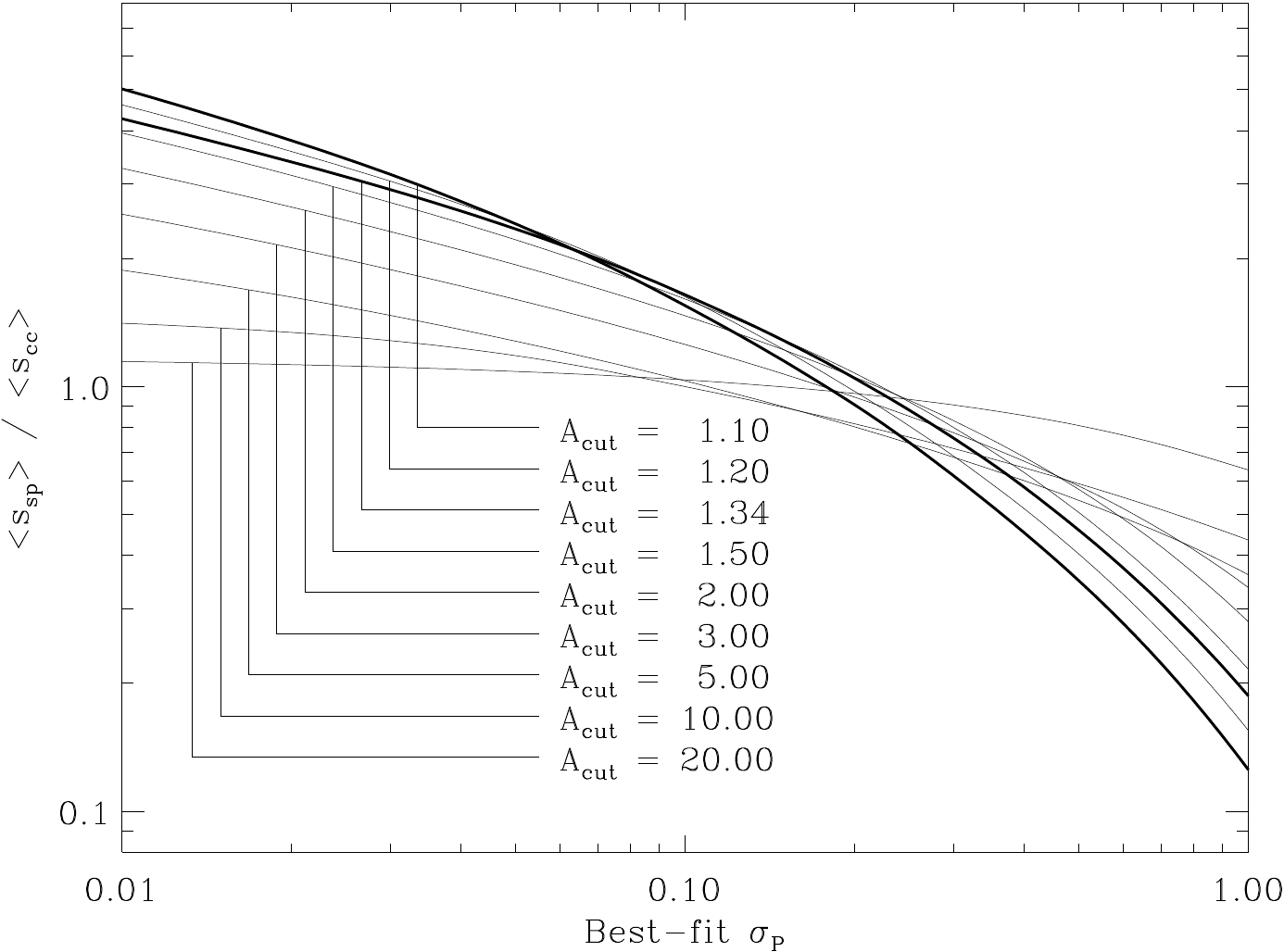} \\
\end{array} $
\end{center}
\caption{Event rate ratio for smoothly perturbed to caustic crossing
events, but considering a variety of event detection thresholds.
Starting with the top curve on the left and proceeding down are values
of $A_{\rm cut}$ of ${1.1, 1.2, 1.34, 1.5, 2, 3, 5, 10}$. The line for
$A_{\rm cut} = 1.1$ shows the same data as Figure
\ref{sigma}.  }
\label{cut}
\end{figure}

\begin{figure}[th]
\begin{center}
$ \begin{array}{c}
\includegraphics[width=3in]{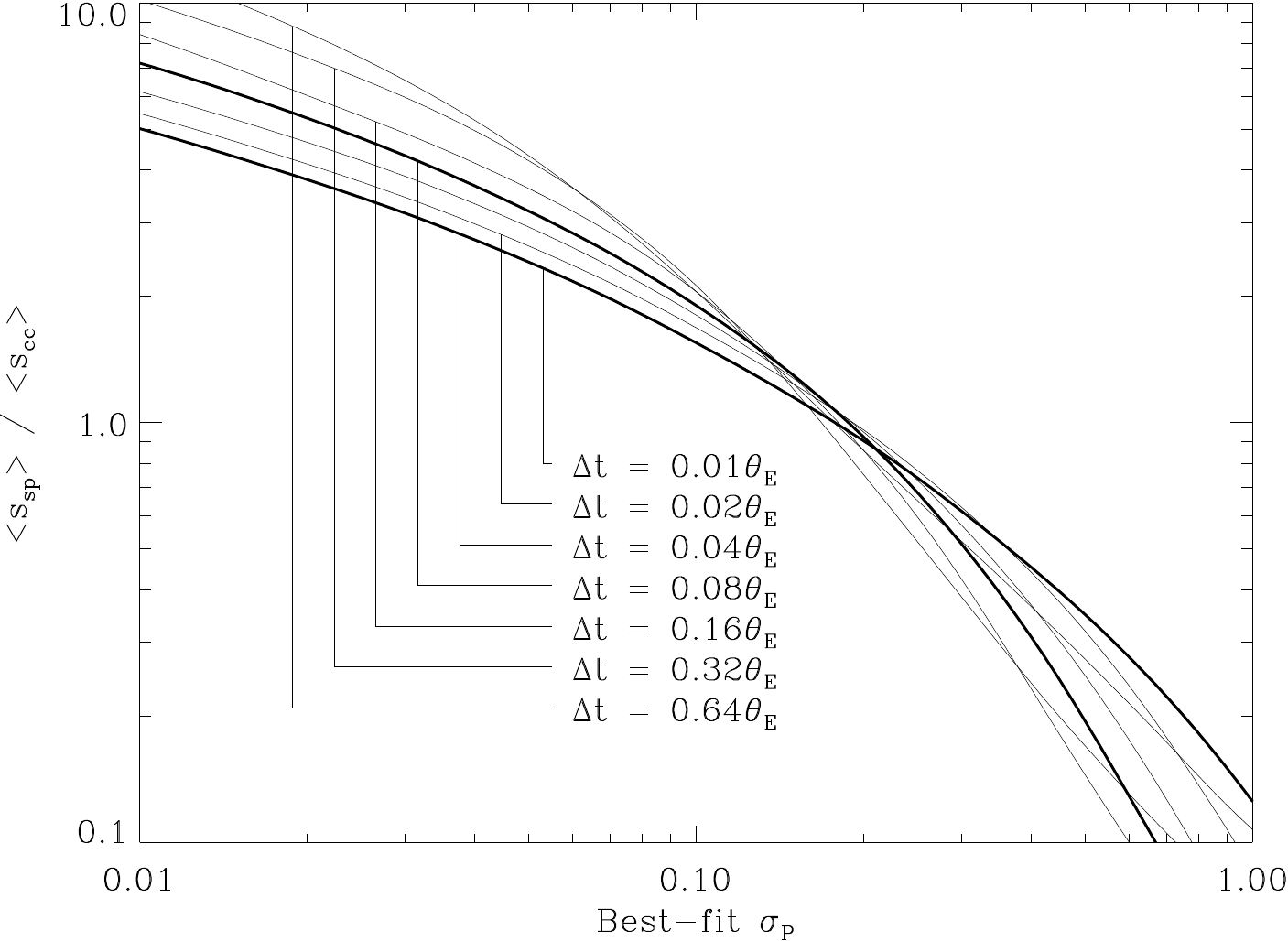} \\
\end{array} $
\end{center}
\caption{Event rate ratio for smoothly perturbed to caustic crossing
events, but considering a variety of light curve sampling rates. The
curve for the default rate of $\Delta t = 0.01 \theta_E$ (showing the
same data as Figure
\ref{sigma}) is on bottom on the left of the plot, and each curve
going up increases $\Delta t$ by a factor of 2.}
\label{sampling}
\end{figure}

\subsection{Robustness of results}
\label{robustness}

Ultimately, we are interested in the relative rates of smoothly
perturbed and caustic crossing events.  It is therefore important to
note that, although we do not include the effects of binary lens
rotation, parallax, binary source rotation, or finite source size,
they all occur in nature. Each of these effects can significantly
change the shape of the light curve, and they would preferentially
perturb light curves away from rather than toward the point lens
form. Only rarely, however, will they change the number of caustic
crossing light curves.  By ignoring these effects, we potentially
compute values of the relative rates of smoothly perturbed events to
caustic crossing events that are lower than the values that should be
observed. Thus, if anything, our results are a lower limit on the
relative rates of smoothly perturbed to caustic crossing events.

A binary source or an extended source would have a more complicated
form, but we assume these complications to be negligibile. Only rarely
is it necessary to invoke a binary source model to fit a binary lens
event, and finite source sizes only significantly affect the shape of
caustic crossings, without changing whether the event is caustic
crossing or not.

\begin{figure}[th]
\begin{center}
\includegraphics[width=3in]{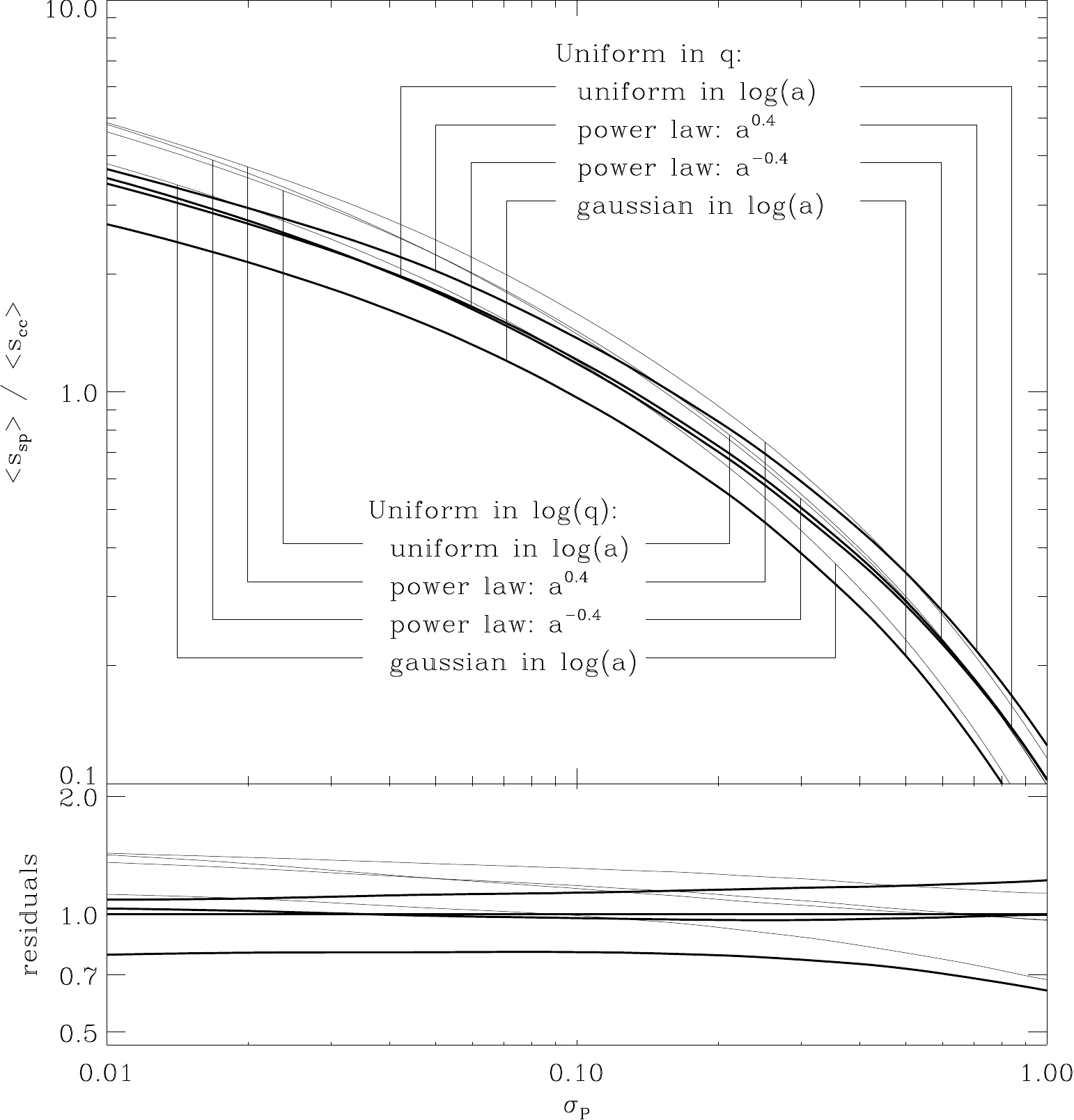} \\
\end{center}
\caption{
Event rate ratio of smoothly perturbed to caustic crossing events,
considering eight different binary lens distribution functions
$P(q,a)$. There are two distributions of $q$ considered, uniform and
log-uniform, and four different distributions of $a$ considered,
uniform, increasing power law ($a^{0.4}$), decreasing power law
($a^{-0.4}$), and Gaussian in log space (centered at $\log(a)=0$ with
a standard deviation of 1 dex). Each of the eight $P(q,a)$ is a
product of two of these. The curves are largely indistinguishable, as
shown in the bottom panel, which is the residuals of the top panel
divided by the default curve of uniform in $q$ and $\log(a)$. The
lowermost curve in this panel represents the distribution uniform in
$q$ and Gaussian in $\log(a)$. }
\label{distribution}
\end{figure}

\begin{figure}[th]
\begin{center}
\includegraphics[width=3in]{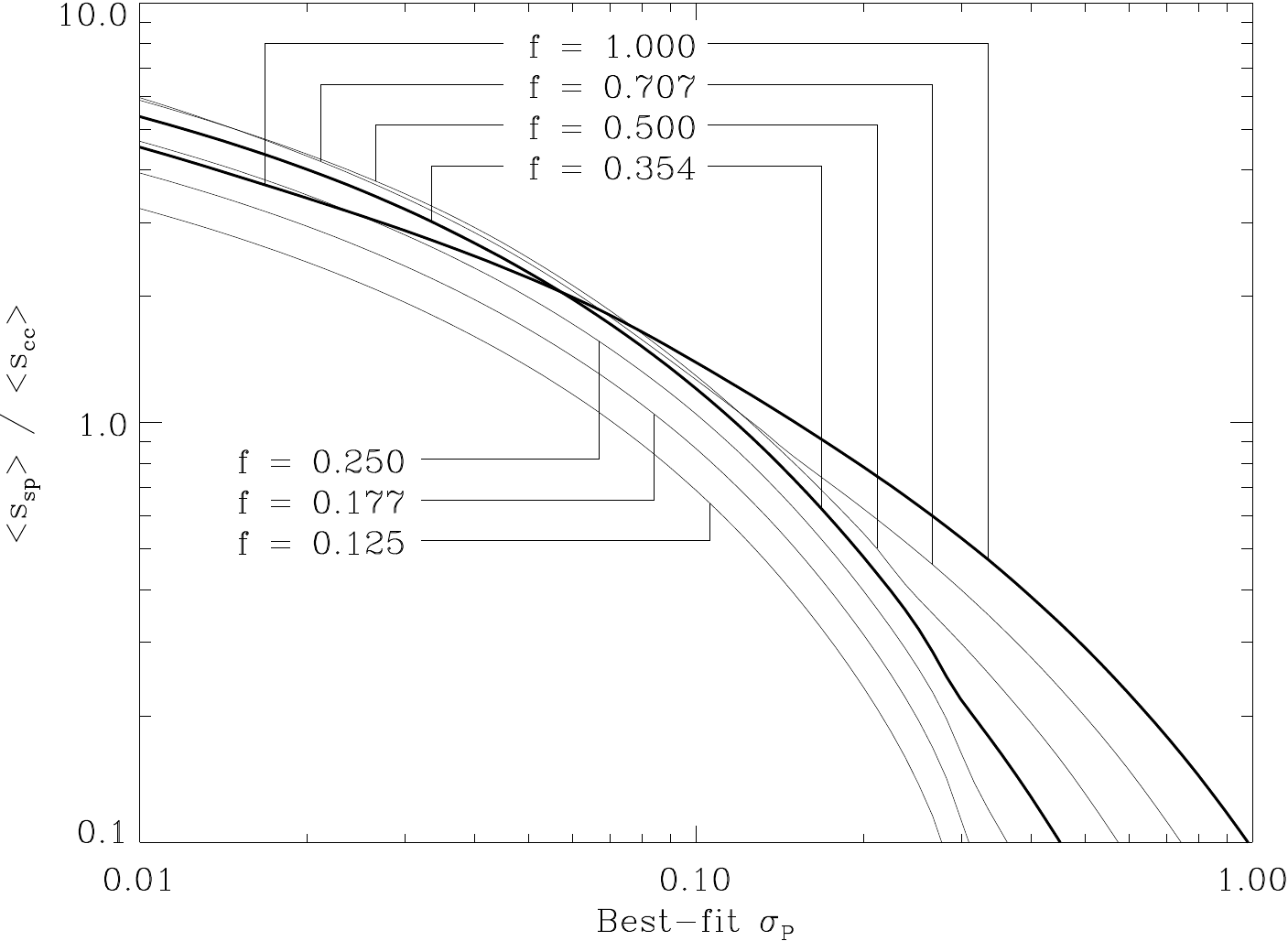} \\
\end{center}
\caption{
Event rate ratio of smoothly perturbed to caustic crossing events,
considering a variety of blending parameters. Each curve corresponds
to a different value of $f$ as shown in the plot, varying from $f =
1.0$ to $f = 0.125$, with consecutive values separated by a factor of
0.707. Two curves, corresponding to $f = 1.0$ and $f = 0.354$, are
shown with thicker lines for contrast. Although there is some overlap
at small values of $\sigma_P$, for $\sigma_P > 0.1$, decreasing $f$
(increasing the blending) systematically decreases the event rate
ratio, as expected. }
\label{blending}
\end{figure}

Our simulation involved several assumptions, in the form of default
parameter values, that were better than a realistic modern
survey. Could one of these assumptions cause our simulations to
produce unrealistic rate ratios? Here we consider variations of the
four most significant assumptions by changing certain parameter
values. Figure \ref{sigma}, showing the rate ratio versus the cutoff
parameter value, as modified by these variations, is shown in a
separate figure for each one. These variations are: event detection
cutoff $A_{\rm cut}$ (Figure \ref{cut}), sampling rate (Figure
\ref{sampling}), binary population distribution function (Figure
\ref{distribution}), and blending parameter (Figure
\ref{blending}).

Decreasing the sampling rate (Figure \ref{sampling}) and considering a
different distribution of binary lenses from log-normal (Figure
\ref{distribution}) will both affect the number of caustic crossing
light curves detected as well as the number of perturbed non-caustic
crossing light curves. Since we can only consider ratios of event
types detected rather than absolute numbers, to evaluate the effects
of these changes we must divide by the new rate of perturbed events by
the new rate of caustic crossing events.

Increasing the event detection threshold $A_{\rm min}$ (Figure
\ref{cut}) and decreasing the blending parameter (Figure
\ref{blending}), on the other hand, selectively affect non-caustic
crossing light curves. These both generally tend to weaken
perturbations and decrease $\sigma_P$ for a given light curve. There
are, however, some exceptions, such as a light curve that is more
similar to a point lens on its wings than at its peak.  Such light
curves will actually produce worse least squares fits when the wings
are cut off by blending or increasing the detection threshold, thus
potentially changing their classification from point lens like to
smoothly perturbed.

Figure \ref{cut} shows various event detection thresholds. $A_{\rm cut} =
1.34$ is often used because for point lens light curves it corresponds
to $b = 1$. Other values to consider are $A_{\rm cut} = 1.46$,
corresponding to $b = 0.75$, and $A_{\rm cut} = 2.03$, corresponding
to $b = 0.3$. Only at very high thresholds, well above these values,
does the event rate ratio become significantly different for $\sigma_P
= 10\%$.

Figure \ref{sampling} shows what happens with a varying sampling
rate. Again, at $\sigma_P = 10\%$, the ratio changes very little for
several factors of 2 away from our default assumption.

Figure \ref{distribution} shows variations due to various binary
lensing populations. Since being a binary has no effect for values of
$a$ very far from unity, only the distribution in $a$ near unity
matters for our rate ratio. So for instance, although a log-normal
distribution in $a$ could not technically extend to infinity, for our
simulation it does not matter where the lower and upper cutoffs are,
as long as they include the orders of magnitude near unity. In
addition to the log-normal distribution in $a$, we test two power law
distributions, with power law exponents of 0.4 and -0.4. Generally,
these have little to no effect on the rate ratios we predict. We also
test a distribution designed to maximize caustic crossings, namely a
distribution Gaussian in $\log(a)$, centered at $\log(a) = 0$, with a
standard deviation of 1.  Even this unrealistic distribution (shown in
the bottom curve of Figure
\ref{distribution}) only lowers the rate ratio by a factor of about 0.8.

Finally, Figure \ref{blending} shows blending considerations. There is
a significant difference in the rates when blendings is taken into
account. Even for a particularly low blending parameter of $f = 0.250$
(where $f$ is the fraction of baseline light from the lensed star),
however, there are still expected to be as many non-caustic crossing
light curves with $\sigma_P > 10\%$ as caustic crossing curves. For
any surveys that are able to counteract the effects of blending
through image differencing techniques, this consideration should not
be applicable, but it may explain a significant part of the
discrepancy for MACHO.

\begin{figure}[ht]
\begin{center}
$ \begin{array}{c}
\includegraphics[width=3in]{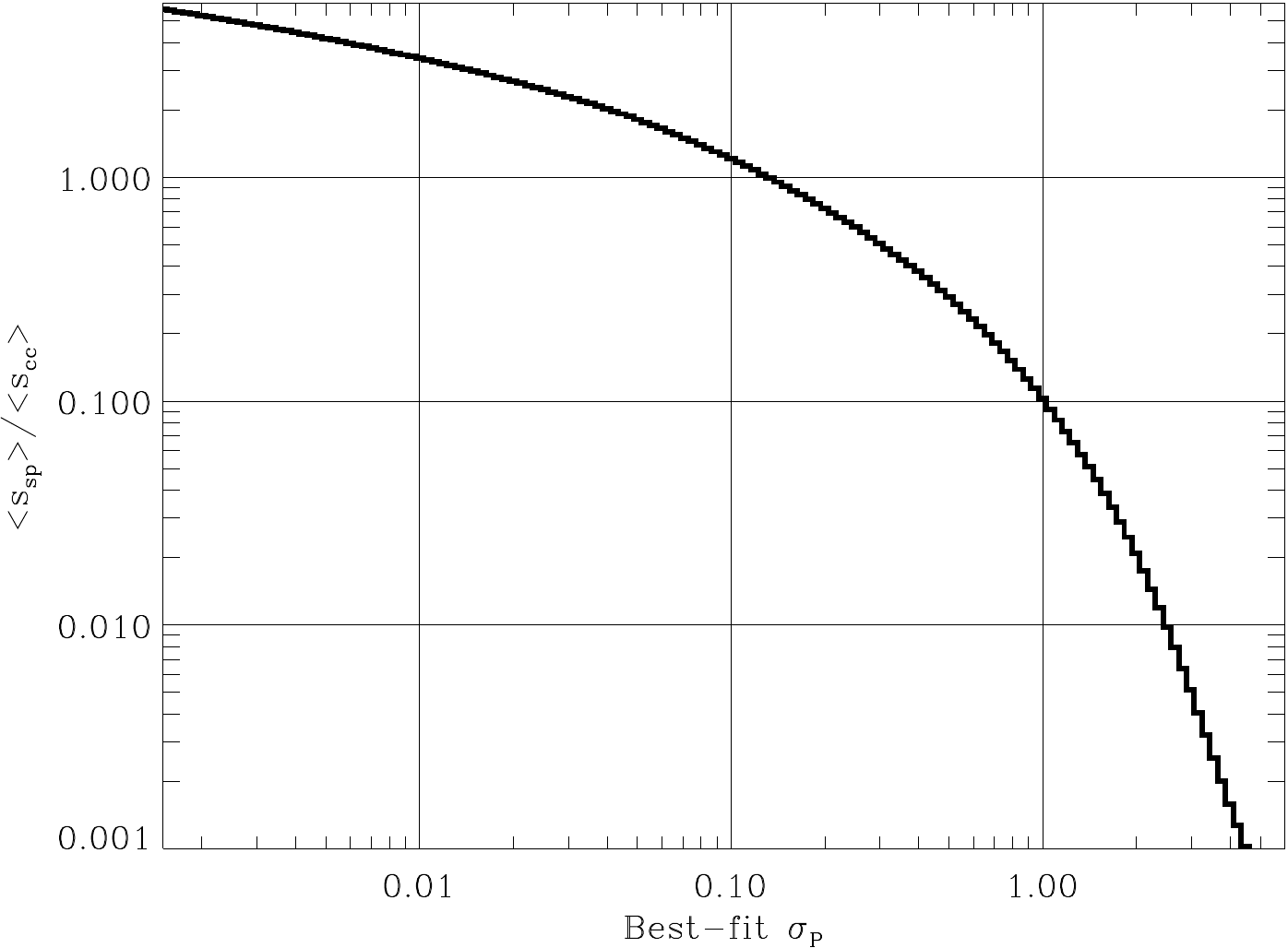} \\
\includegraphics[width=3in]{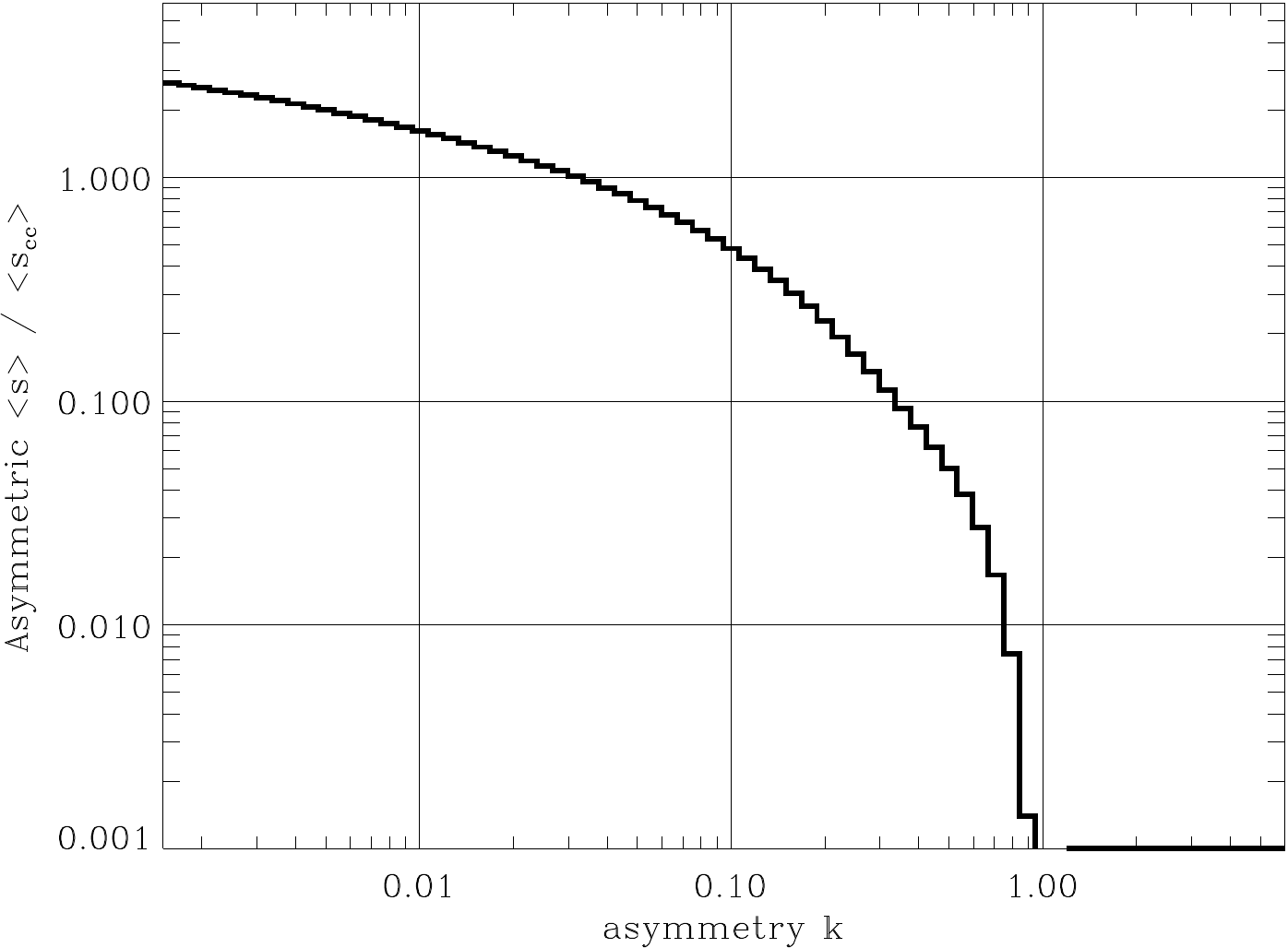} \\
\includegraphics[width=3in]{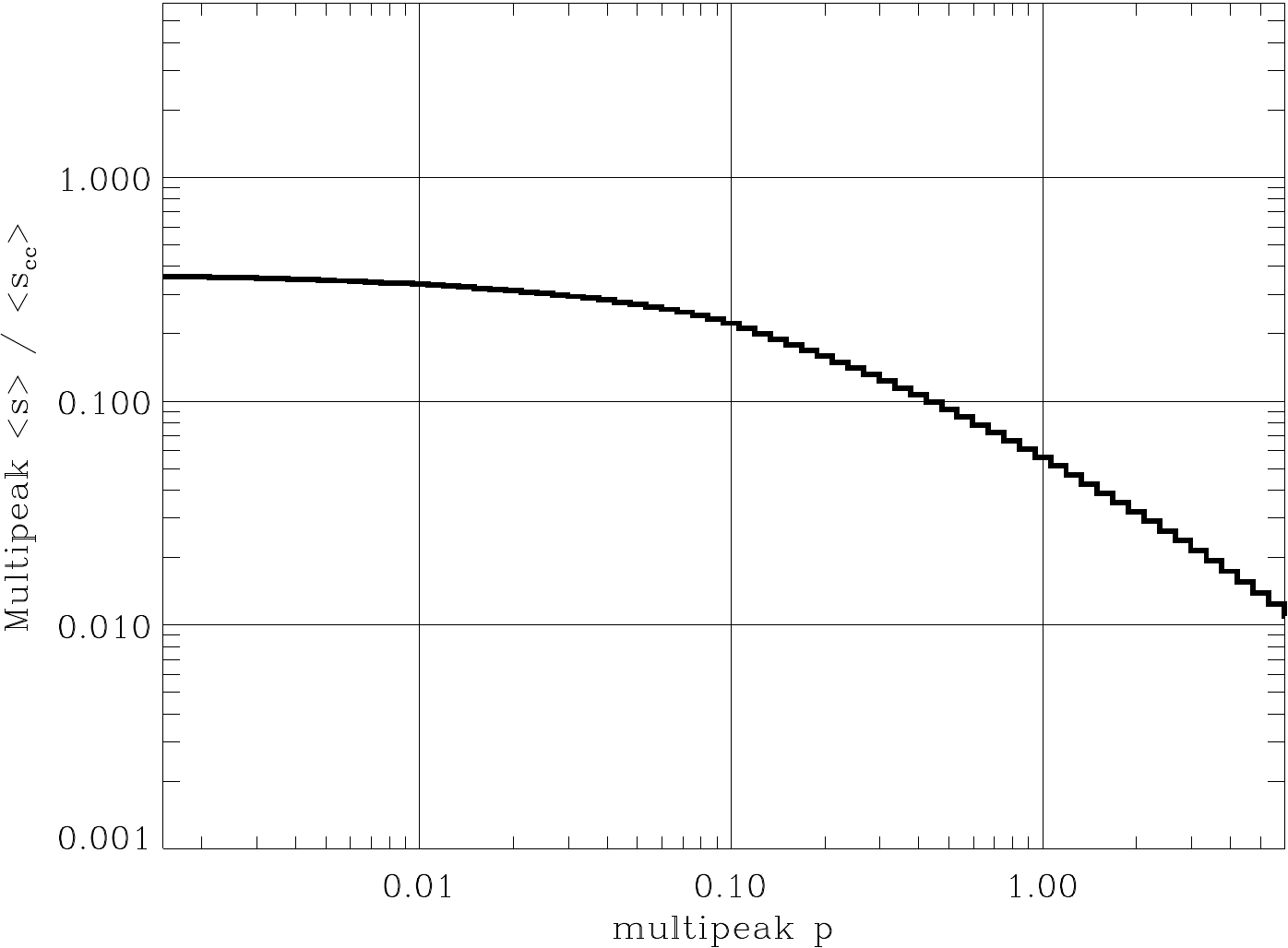} \\
\end{array} $
\end{center}
\caption{ Histograms showing the cumulative distribution function
for the three parameters for non-repeating, non-caustic crossing light
curves. For any particular value of a parameter, the graph shows the
value $\langle s\rangle$ for all light curves with at least the given
value. The data in the top panel are identical to Figure \ref{sigma}.
}
\label{parhist}
\end{figure}

\begin{figure*}[ht]
\begin{center}
\includegraphics[width=6in]{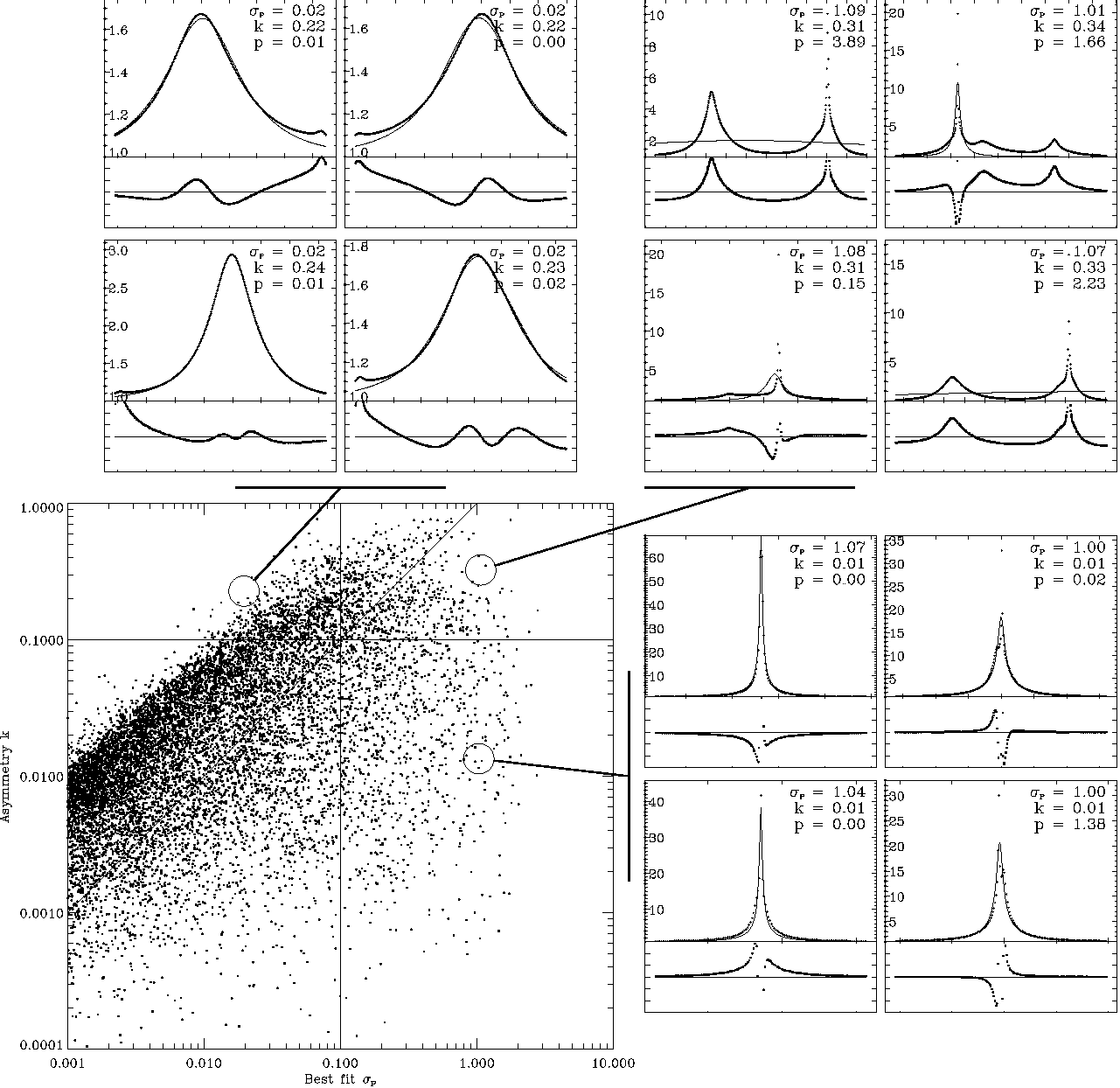}
\end{center}
\caption{ Scatter plot showing the relative distribution of the best-fit
parameter $\sigma_P$ and the asymmetry parameter $k$. Horizontal and
vertical lines denote parameter values of 0.10, and along the diagonal
line the two parameters are equal. Note that there are a significant
number of light curves for which one parameter is greater than 0.10
but not the other. Sample light curves are shown to represent three
different places in the parameter space: above are four curves with
significant asymmetry but good point lens fits; right are four
symmetric curves without good point lens fits; above and right are
four curves with high values of both parameters. Beneath each light
curve is shown the residual plot for the best point lens fit. For clarity,
this scatter plot shows only one in every 10 points compared with the
next two plots. }
\label{scatter1}
\end{figure*}

\begin{figure*}[ht]
\begin{center}
\includegraphics[width=6in]{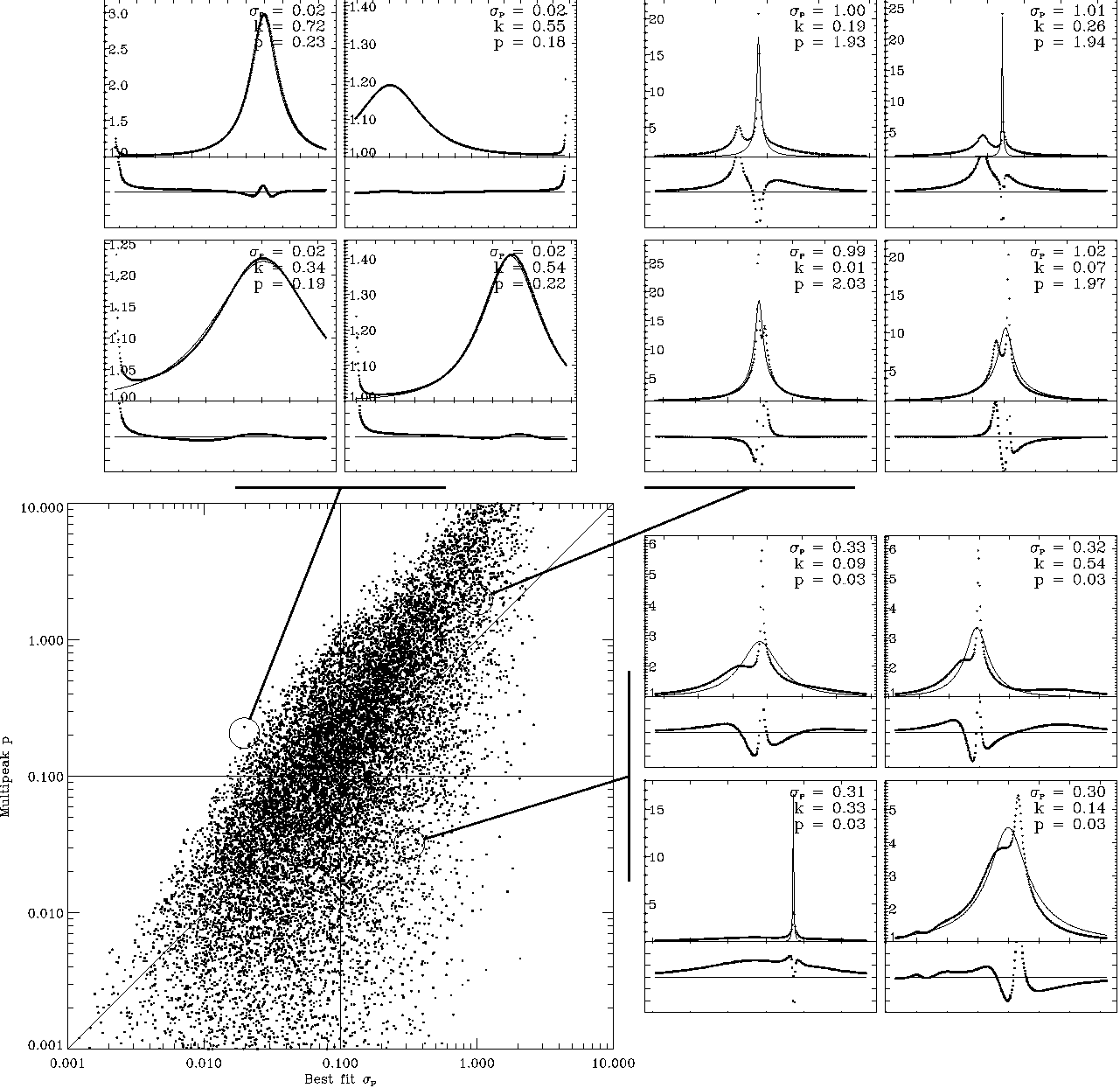}
\end{center}
\caption{ Scatter plot showing the relative distribution of the best-fit
parameter $\sigma_P$ and multipeak parameter $p$. Horizontal and
vertical lines denote parameter values of 0.10, and along the diagonal
line the two parameters are equal. Again, there are a significant
number of light curves for which one parameter is greater than 0.10
but not the other. Sample light curves are shown to represent three
different places in the parameter space: above are four curves with
significant secondary peaks but good point lens fits; right are four
curves with only very small secondary peaks, but without good
point lens fits; above and right are four curves with high values of
both parameters. Beneath each light curve is shown the residual plot
for the best point lens fit. Single-peaked events have $p = 0$ and so
are not represented in this scatter plot. }
\label{scatter2}
\end{figure*}

\begin{figure*}[ht]
\begin{center}
\includegraphics[width=6in]{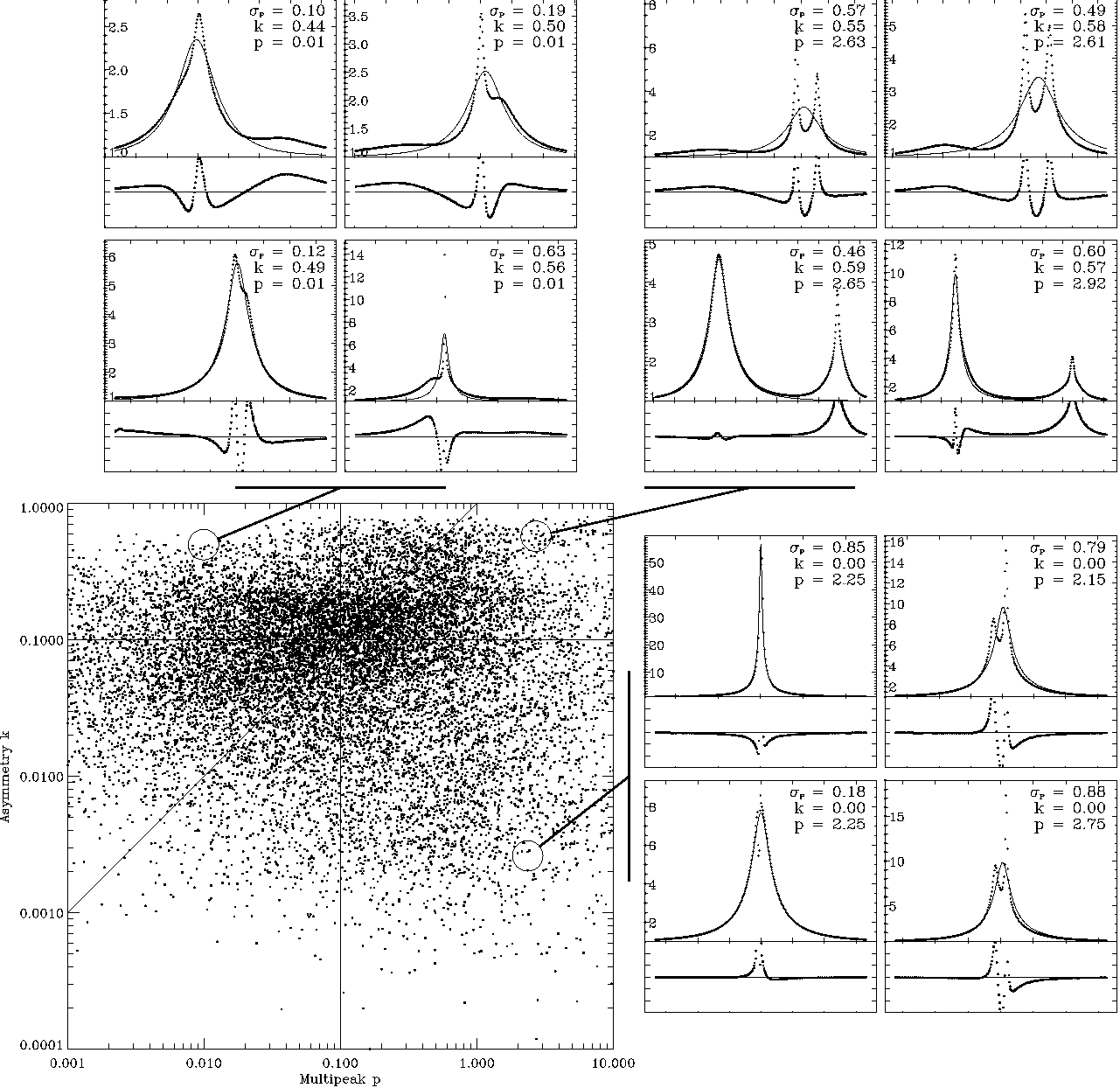}
\end{center}
\caption{ Scatter plot showing the relative distribution of the multipeak
parameter $p$ and the asymmetry parameter $p$. Horizontal and vertical
lines denote parameter values of 0.10, and along the diagonal line the
two parameters are equal. Again, there are a significant number of
light curves for which one parameter is greater than 0.10 but not the
other. Sample light curves are shown to represent three different
places in the parameter space: above are four curves with very small
secondary peaks but high asymmetry; right are four symmetric curves
with significant secondary peaks; above and right are four curves with
high values of both parameters. Beneath each light curve is shown the
residual plot for the best point lens fit. Single-peaked events have
$p = 0$ and so are not represented in this scatter plot. }
\label{scatter3}
\end{figure*}

\subsection{Additional ways to detect perturbations}

In the work presented above we have used the value of $\sigma_P$ to
determine whether a light curve should be viewed as smoothly perturbed
from the point lens form.  Other parameters, such as the asymmetry
parameter, $k$, and the multipeak parameter, $p$ have been defined and
calculated to provide ways of characterizing deviations from the
point lens form.  It is important to note, however, that by using
$\sigma_P,$ we are employing a measure of the deviation of the light
curve as a whole. Light curves with short-lived deviations from the
point lens form may not be characterized as smoothly perturbed for
values of $\sigma_P \sim 0.1,$ even if current observational setups
can detect the short-lived deviations at a high level of significance.

In other words, by using $\sigma_P$ to identify smoothly perturbed
light curves, we are taking a conservative approach.  This seems
appropriate, given that it has so far been relatively difficult to
identify smoothly perturbed light curves in real data
sets. Nevertheless, in future work it may be appropriate to extend the
definition of smoothly perturbed light curves to include some with
small values of $\sigma_P,$ but large values of $k$ or $p.$

We therefore consider the distribution of the three parameters
$\sigma_P$, $k$, and $p$ that is produced by the default binary
distribution in \S \ref{population-results}. We consider the
parameters independently (Figure \ref{parhist}) and also in relation
to one another (Figures \ref{scatter1} through \ref{scatter3}). These
figures suggest that it would be worthwhile to augment least squares
fitting by considering other avenues of identifying binary lens
events. If this is done, some events that would otherwise be
classified as point lens like will instead be classified as smoothly
perturbed. The predicted values of $\langle s_{sp} \rangle/\langle
s_{cc} \rangle$ would then be even larger.
    
\section{Conclusions}
\label{conclusions}

We have explored and categorized the full range of microlensing light
curves produced by binary lenses.  The three mutually exclusive
categories of binary lens light curves we have studied are point lens
like, smoothly perturbed, and caustic crossing.  Smoothly perturbed
light curves are defined to be those that exhibit continuous
deviations from the point lens form; this naturally excludes caustic
crossings.  In this paper, smoothly perturbed light curves have been
identified by their failure to be fit by a point lens model with a
least squares metric.  We have also determined whether each light
curve we have computed exhibits asymmetry with respect to time
reversal, and/or multiple peaks. For the purposes of this paper, we
have used an asymmetry parameter and a multipeak parameter only to
characterize the deviations from the point lens form, not to identify
smoothly perturbed light curves.

Because we have sampled a wide range of binary separations, we have
also covered repeating events \citep{DiStefanoMao}. Our simulation
results are consistent with the analytically calculated rates
\citep{DiStefanoScalzo2} and predict that repeating events will form
a significant part of data sets that are sensitive to deviations at
the few percent level. The rate of repeating events increases with the
size of the lensing region, and so it increasing significantly with
improved photometric precision. The sensitivity of existing data sets
should be enough for repeating events to constitute a few percent of
binary lens events, so as the number of binary events approaches 100
and sensitivity continues to improve, repeating events become
inevitable.

As with repeating events, other types of exotic events that are
expected to constitute just a few percent of binary lens events will
also become inevitable with growing data sets.

\subsection{The missing smoothly perturbed light curves} 

Our most striking result relates to the relative number of caustic
crossing events. We find that under most assumptions about the binary
lens population and about the observational sampling, there should be
more smoothly perturbed events than caustic crossing events. This is
in contrast to the observational results to date, in which binary lens
events are dominated by caustic crossing events.

To determine the reasons for this discrepancy, we have conducted a
range of simulations. We find that the ratio is not very sensitive to
changes in the binary population, specifically the distribution of
binary parameters. The reason for the discrepancy must therefore be
related to the observational setup or to the analysis. We also find,
however, that the ratio is not very sensitive to changes in the cutoff
magnification or sampling frequency. It is of course sensitive to
changes in the photometric uncertainty, but it would have to be much
larger than $10\%$ to account for the discrepancy. Note that we have
considered only the case in which the fractional photometric
uncertainties are constant.  On the other hand, the observational
uncertainties near peaks in magnification may tend to be smaller. If
so, for a given value of the uncertainty at baseline, more deviations
from the point lens form could be observed.  This would further
increase the fraction of events that are recognized as being smoothly
perturbed.

Is it possible that some of the physical effects we have neglected in
our simulations could produce more caustic crossing events and fewer
smoothly perturbed events? The effects we have neglected include
binary rotation, finite source size effects, binary sources, and
parallax.  All of these effects can produce perturbations from the
point lens form.  Only rotation is likely to lead to more caustic
crossings, and this is expected to happen only rarely. Therefore,
calculations that include all of the expected physical effects will
likely produce even larger rate ratios.

Only one physical effect seems to have a significant influence, and
that is blending. Blending cannot obscure the wall-like features that
mark caustic crossings, but it can smooth out the distinctive features
of a smoothly perturbed event, making it more likely that a point lens
fit will be successful. Blending is therefore expected to decrease the
ratio $\langle s_{sp} \rangle/\langle s_{cc} \rangle$.  Nevertheless,
blending would have to be severe, with only $\sim 12\%$ of the
baseline light coming from the lens star, in order for blending to be
responsible for the discrepancy between the predictions and the
results derived so far.  Furthermore, because the present generation
of monitoring programs uses image differencing, which mitigates the
effects of blending, this cannot explain why recent results still
seem to favor caustic crossings \citep{OGLEBinaries2004}.

\subsection{Searching for the missing events}

The discussion above indicates that the lensing programs are likely to
have detected smoothly perturbed binary lens light curves, which were
not identified as such.  Instead, some smoothly perturbed events may
have been identified and published as point lens light curves. If
this were the case for all smoothly perturbed light curves, then the
experimentally derived lensing event rate would be correct, but it
would be difficult to assess the contributions of binary lenses.

On the other hand, the perturbations of some smooth binary lens light
curves may be so pronounced that they have prevented the events from
being identified as lensing candidates at all
\citep{DiStefanoPerna}. This was especially likely during the early
phase of microlensing studies, when it was important for selection
criteria to be conservative, to be certain that the selected events
were truly associated with microlensing.

There are several important reasons to attempt to identify all binary
lens events. The first is to make reliable measurements of the optical
depth. Beyond the direct considerations of determining the rate of
events, the interpretation of the events can help to determine the
locations of significant populations of lenses. Already several binary
lenses in the direction of the Magellanic clouds have been located by
the caustic crossing events they caused. As these events with
measurable lens distances are only a fraction of all events caused by
the lens population, a large number of all lenses can be located
indirectly. \citet{DiStefano2000} used this argument to demonstrate
the possibility that most of the Magellanic Cloud events were caused
by self-lensing. To confirm or refute this argument, we must be able
to identify the smoothly perturbed binary lens light curves.

The second reason it is important to identify all binary lens events
is to learn about the characteristics of the binary lens and
planet lens populations. One important strength of microlensing planet
searches is that, of all methods of planet and binary detection,
microlensing alone can identify the locations and measure some of the
properties of planets in distant stellar systems such as external
galaxies. To conduct a genuine population study, however, we must be
able to identify a wide range of events, and to understand the
detection efficiencies for each type of event.

\subsection{Future monitoring programs} 

The ability of monitoring programs to detect lensing events has
increased dramatically since the first published events. The OGLE team
now routinely identifies roughly 500 lensing events per year, compared
with a total of 9 events identified in its first two years
\citep{OGLE1994}. New monitoring projects will be focused on
microlensing, while wide field monitoring conducted by Pan-STARRS
\citep{PanSTARRS} and
LSST \citep{LSST} will identify thousands of events per year caused
by lenses in the source galaxies, nearby lenses, and also by MACHOs,
should they exist \citep{DiStefano2007b}. The photometric sensitivity
of these projects will approach the level of $1\%$. As illustrated in
Figure \ref{sigma}, this means that these programs should be able to
identify about three times as many smoothly perturbed binary lens
light curves as caustic crossing light curves, using least squares
fitting alone. If, in addition, pronounced asymmetries or correlated
residuals can be used to quantify the probability that some light
curves with acceptable point lens fits were caused by binaries,
smoothly perturbed light curves may play an even larger role in the
data sets.

In order for future monitoring projects to place constraints on the
form of dark matter in MACHOs and to study the underlying
characteristics of each lens population, including planet lenses,
methods to detect the full range of binary lens and planet light
curves must be developed, and the relevant detection efficiencies must
be quantified.

\vspace{+0.5cm}
\noindent Acknowledgements: 
RD would like to thank Rosalba Perna and Nada Petrovic for
conversations. Funded in part by NASA NAG5-10705, PHY05-51164, and a
grant for SAO Internal Research \& Development.

\end{document}